\documentclass[12pt]{article}
\PassOptionsToPackage{shortlabels}{enumitem}
\usepackage[utf8]{inputenc}
\usepackage[T1]{fontenc}
\usepackage{amsmath,amssymb,amsthm,commath,enumitem,mathtools}
\usepackage{bm}
\usepackage{kbordermatrix}
\usepackage[singlelinecheck=false]{caption}
\usepackage[margin=2.54cm]{geometry}
\usepackage[normalem]{ulem}
\usepackage{natbib}
\usepackage{graphicx}
\usepackage{booktabs}
\usepackage{color} 
\usepackage[hyphens]{url} 
\usepackage[hidelinks]{hyperref}
\usepackage[onehalfspacing]{setspace}
\usepackage[small]{titlesec}
\usepackage{enumitem}
\usepackage[ruled,vlined]{algorithm2e}
\usepackage[tableposition=top,labelfont=bf]{caption}
\usepackage{layouts}
\usepackage{xcolor}
\usepackage{makecell}
\usepackage{tabularray}
\graphicspath{ {./figures/} }
\usepackage{fancyhdr}

\allowdisplaybreaks
\let\originalleft\left
\let\originalright\right
\renewcommand{\left}{\mathopen{}\mathclose\bgroup\originalleft}
\renewcommand{\right}{\aftergroup\egroup\originalright}
\makeatletter
\newcommand*\bigcdot{\mathpalette\bigcdot@{.5}}
\newcommand*\bigcdot@[2]{\mathbin{\vcenter{\hbox{\scalebox{#2}{$\m@th#1\bullet$}}}}}
\makeatother

\title{%
   {Hidden Markov Individual-level Models of Infectious Disease Transmission} \\[5pt]
  }
\author{Dirk Douwes-Schultz$^{1}$\footnote{{{\it Corresponding author}: Dirk Douwes-Schultz, Department of Mathematics and Statistics, University of Montreal, 2920 Tour Rd, Room 4225, Montreal, QC, Canada H3T 1N8. \\ {\it E-mail}: {\tt
				{\color{blue}dirk.douwes-schultz@umontreal.ca}}.}} , Rob Deardon$^{1,\color{black}2}$ and Alexandra M. Schmidt$^{3}$ \\
                	$^{1}$\textit{Department of Mathematics and Statistics,}  \textit{University of Calgary, Canada } \\
                    $^{\color{black}2}$\textit{\color{black}Faculty of Veterinary Medicine,} \textit{\color{black}University of Calgary, Canada } \\
				$^{3}$\textit{Department of Epidemiology, Biostatistics and Occupational Health,} \\ \textit{McGill University, Canada }}
				
\date{\today}

\begin{document}

\maketitle

\begin{abstract}

Individual-level epidemic models are increasingly being used to help understand the transmission dynamics of various infectious diseases. However, fitting such models to individual-level epidemic data is challenging, as we often only know when an individual {\color{black}first showed symptoms of the disease} and not when they were infected or removed. We propose an autoregressive coupled hidden Markov model to infer unknown infection and removal times, as well as other model parameters, {\color{black}given only the symptom-onset times}.  {\color{black}More traditional data augmentation methods used in epidemic modelling have either assumed that the infection or removal times are known or that all infected individuals showed symptoms. In contrast, we build an observation model where there is a chance that an infectious individual never shows symptoms during the study, allowing for undetected (e.g., asymptomatic) infections. Symptom onset is not assumed to correspond to infection or removal.}  Bayesian coupled hidden Markov models have been used previously for individual-level epidemic data {\color{black} consisting of laboratory diagnostic test results}. However, these {\color{black} models are not appropriate for symptom-onset data, as they have a chance of generating multiple symptom-onset times for an individual. In contrast, we incorporate autoregression into the observation process, ensuring that at most one symptom-onset time is observed per individual.} We illustrate the flexibility of our approach by fitting two examples: an experiment on the spread of tomato spot wilt virus in pepper plants and an outbreak of norovirus among nurses in a hospital.

\end{abstract}

\section{Introduction} \label{section:intro}

Epidemiologists are often interested in questions related to the transmission of an infectious disease at the individual level. For example, whether susceptibility differs by individual-level characteristics such as age \citep{cohen1997social,davies2020age}, or how far an infected individual could realistically spread the disease  \citep{hu2021risk,lichtemberg2022dispersal}. Individual-level models (ILMs) of infectious disease transmission \citep{deardon_inference_2010} can be valuable tools for helping answer these types of questions \citep{vynnycky2010introduction}. {\color{black}These} approaches model each individual in the population moving through different disease states, {\color{black}such as} susceptible, infectious, and removed \citep{ward2025framework}. Transitions between states can occur in continuous \citep{almutiry2021continuous} or discrete \citep{v_individual-level_2020} time. We will focus on discrete time. For discrete-time models, the probability of infection at each time step may depend on the number of infectious individuals in the population, their distance from the susceptible individual, and the inherent susceptibility or infectivity of individuals, which may vary with covariates \citep{keeling2001dynamics,mahsin_geographically_2022}. Therefore, these models can describe a wide range of complex mixing patterns.

However, a significant challenge with fitting ILMs to individual-level epidemic data is
that we usually only know when {\color{black}symptoms first appeared in an individual (i.e., their symptom-onset time)}, not when they were infected or
removed \citep{neal_statistical_2004,kypraios_tutorial_2017}. For example, in Section \ref{sec:TSWV}, we look at an experiment
on the spread of tomato spotted wilt virus (TSWV) in pepper plants, which was described in
\cite{hughes1997validating} and analyzed using ILMs by \cite{almutiry2021continuous}. The data consists of the symptom-onset time of each plant that showed symptoms during the experiment. Since symptoms of TSWV take 2-4 weeks to appear in a plant, {\color{black} we do not know when a plant that showed symptoms was infected (there is also no information on when/if a plant was removed). Furthermore, just because a plant never showed symptoms during the experiment does not mean it was not infected. Some plants that did not show symptoms may have had asymptomatic infections or were just infected late in the experiment and had not yet developed symptoms (it takes 2-4 weeks for symptoms to develop, so any plant infected in the last week or so of the experiment would likely not have shown symptoms). Therefore, we do not know the infection or removal times or even how many plants were infected, which makes inference challenging.} 

The most popular way to account for uncertain infection and removal times in epidemic modelling is to treat them as unknown parameters of the model within a Bayesian framework, which is known as data augmentation (DA) \citep{o2002tutorial,o2019markov}.  Most DA methods assume that infection times are unknown and removal times are known \citep{oneill_bayesian_1999,jewell_bayesian_2009,deardon_inference_2010}, or that infection times are known and removal times are unknown \citep{bu_likelihood-based_2022}. However, in many applications, we do not observe when individuals were removed or infected \citep{neal_statistical_2004,pokharel2022emulation}. For such cases, \cite{almutiry2021continuous} proposed using a susceptible-infectious-notified-removed (SINR) model \citep{jewell_bayesian_2009} to estimate both unknown infection and removal times using observed notification ({\color{black}usually symptom-onset}) times.  However, they assumed that only those who showed symptoms during the study were infected (i.e., that there were no undetected infections). {\color{black} Some of the individuals who did not show symptoms may have had asymptomatic infections or been infected near the end of the observation period and had not yet developed symptoms \citep{jewell_bayesian_2009}. Under \cite{almutiry2021continuous}, these individuals will incorrectly be assumed to have escaped infection, which can lead to an underestimation of the transmission intensity (we show this in Section \ref{sec:TSWV}).}  \cite{cauchemez_bayesian_2004} and \cite{angevaare_inference_2021} used similar methods to \cite{almutiry2021continuous}, but again did not account for undetected infections.

Taking a {\color{black}very} different approach, \cite{touloupou_scalable_2020} proposed using a coupled hidden Markov model \citep{pohlePrimerCoupledStateswitching2021} to account for unknown infection and removal times for discrete-time ILMs. They first assumed that the epidemiological states of the individuals (e.g., susceptible, infectious, and removed) followed a series of coupled hidden Markov chains. By coupled, we mean that the transition probabilities of an individual chain, like the probability of going from susceptible to infectious, could depend on whether other individuals were in the infectious state {\color{black}to account for disease spread}. Then they generated the {\color{black}observations (e.g., test results)} conditional on the hidden states using an independent Bernoulli observation model. There are many important advantages of this approach over the more traditional DA methods described in the previous paragraph. Firstly, {\color{black}unlike most of those methods,} their approach does not assume {\color{black}that any of the removal or infection times are known. In addition, infectious individuals have a chance of not being detected by investigators during the observation period. Therefore, unlike the SINR approach of \cite{almutiry2021continuous}, they account for the possibility that an undetected individual (e.g., someone who did not show symptoms in our applications) was infected.}

{\color{black}However, \cite{touloupou_scalable_2020} assumed that each individual was given laboratory diagnostic tests for the disease at predetermined times (e.g., twice a week). In epidemiological studies of outbreaks, often only the time symptoms first appeared in an individual is known, and no temporal laboratory testing information is available \citep{neal_statistical_2004,kypraios_tutorial_2017,stockdale_modelling_2017,stockdale_pair-based_2021}. In addition, even if such data is available, there is often a need to supplement it with symptom-onset data to, for example, study asymptomatic infections \citep{tsang2023reconstructing}. As \cite{touloupou_scalable_2020} used an independent Bernoulli observation model, see Equation (\ref{eqn:GHMM_obs_cont}), their approach can be applied to symptom-onset data (i.e., binary data where a 1 denotes symptom onset; see Section \ref{sec:HMMILM}) or any binary observation data. However, their model is not appropriate for such data, as it has a chance of generating multiple symptom-onset times for an individual, as we show mathematically in Section \ref{sec:TSWV}.}

To address the above limitations, we extend the approach of \cite{touloupou_scalable_2020} by incorporating autoregression into the observation process. This allows us to {\color{black}ensure that at most one symptom-onset time is observed per individual.} {\color{black}The} resulting hidden Markov model is very general. {\color{black}It only requires the symptom-onset times, which are easy to observe and often available in outbreak investigations as they form the epidemic curve \citep{thompson_faith_1968,caceres1998viral,aristotelous_posterior_2023}.} Additionally, it does not assume any of the infection or removal times are known, {\color{black}and infectious individuals do not have to show symptoms during the study (i.e., it can generate undetected infections).} 

The remainder of this paper is organized as follows. Section \ref{sec:HMMILM} introduces {\color{black}the proposed} {\color{black}individual-level epidemic hidden Markov model for symptom-onset times}. Section \ref{sec:inf_proc} {\color{black}discusses} our Bayesian inferential procedure, which makes use of the {\color{black}computationally} efficient iFFBS algorithm of \cite{touloupou_scalable_2020}. In Section \ref{sec:applications}, we fit our proposed model to two examples: an outbreak of norovirus in a hospital and an experiment on the spread of tomato spotted wilt virus in pepper plants. We compare {\color{black}our approach} to several popular alternatives from the literature, such as models that assume known infection or removal times or {\color{black}that all infected individuals showed symptoms}. We close with a discussion in Section \ref{sec:discussion}.

\section{Hidden Markov Individual Level {\color{black}Modelling Framework}}
\label{sec:HMMILM}

\subsection{Transmission model}

Assume we have a population of $i=1,\dots,N$ individuals who are observed across $t=1,\dots,T$ time periods. Let $S_{it}$ be an indicator for the true epidemiological state of the individual, where $S_{it}=1$ if individual $i$ was susceptible during time $t$, $S_{it}=2$ if individual i was infectious, and $S_{it}=3$ if individual $i$ was removed. Following \cite{deardon_inference_2010}, we assume that $S_{it}$ follows a three-state nonhomogeneous Markov chain. In order to account for disease spread between individuals, we condition the transition matrix on $\bm{S}_{(-i)(t-1)}=(S_{1(t-1)},\dots,S_{(i-1)(t-1)},S_{(i+1)(t-1)},\dots,S_{N(t-1)})^T$, the vector of disease states for all individuals excluding individual $i$ at time $t-1$. We propose the following conditional transition matrix for the Markov chain, for $t=1,\dots,T$, \renewcommand*{\arraystretch}{1} 
\begin{align} \label{eqn:TMatrix}
&\Gamma(S_{it}|\bm{S}_{(-i)(t-1)})=\kbordermatrix{
\textbf{State}  & S_{it}=1 \, \textbf{(sus.)} & &  S_{it}=2 \, \textbf{(inf.)} & & S_{it}=3 \, \textbf{(rem.)}  \\[8pt]
  S_{i(t-1)}=1 \, \textbf{(sus.)} & 1-p_{12,it}& & p_{12,it} & & 0 \\[5pt]
  S_{i(t-1)}=2 \, \textbf{(inf.)} & 0 & & 1-\frac{1}{m} & & \frac{1}{m} \\[5pt]
    S_{i(t-1)}=3 \, \textbf{(rem.)} & 0 & & 0 & & 1
  },
\end{align}  where $\Gamma(S_{it}|\bm{S}_{(-i)(t-1)})_{lk}=P(S_{it}=k|S_{i(t-1)}=l,\bm{S}_{(-i)(t-1)})$ {\color{black} and $m$ is the average duration of the infectious period.}

In (\ref{eqn:TMatrix}), $p_{12,it}$ represents the probability that individual $i$ is infected in the interval $[t-1,t)$ if they are susceptible at time $t-1$. We will assume individual $i$ primarily makes contacts within a neighbourhood $NE(i)$. Let $\beta_{j \to i,t}$ be the probability of an infectious contact between individuals $i$ and $j \in NE(i)$ in the interval $[t-1,t)$. Also, suppose there is a small constant risk $\alpha$ of an individual being infected by someone outside their neighbourhood (potentially outside the population) or a background reservoir. Then, assuming all contacts are independent, we have 
\begin{align}
p_{12,it} &= 1-(1-\alpha)\prod_{j \in NE(i)}(1-\beta_{j \to i,t})^{I[S_{j(t-1)}=2]} \notag \\[5pt]
& \approx 1-\exp\left(-\alpha-\sum_{j \in NE(i)}\beta_{j \to i,t}I[S_{j(t-1)}=2]\right), \label{eqn:prob_infect}
\end{align} which follows the classic Reed-Frost formulation of a discrete-time SIR model \citep{vynnycky2010introduction,deardon_inference_2010}. Note that the approximation (\ref{eqn:prob_infect}) holds only when $\beta_{j \to i,t}$ is small for all $j \in NE(i)$, which may not be appropriate if some individuals in the population are closely connected. Therefore, we more generally interpret $\beta_{j \to i,t}>0$ as the effect of disease spread from individual $j$ to individual $i$ in the {\color{black}continuous time} interval $[t-1,t)$, and we generally do not restrict it to be less than 1.

{\color{black}\cite{deardon_inference_2010} proposed a very flexible form for the effects of disease spread, $\beta_{j \to i,t}=\mathcal{S}_i \mathcal{T}_j  \mathcal{K}_{ij}$, where  $\mathcal{S}_i > 0$ is a susceptibility function for susceptible individual $i$, $\mathcal{T}_j > 0$ is a transmissibility function for infective $j$, and $\mathcal{K}_{ij} > 0$ is an infection kernel representing some contact measure between $i$ and $j$ (e.g., distance). This form has been extended in a myriad of ways, for example, to allow for geographically varying parameters \citep{mahsin_geographically_2022} and to account for covariate measurement error \citep{amiri2024spatial}. For ease of exposition, and since we mainly focus on the observation model, we assume a more generic specification for the effect of disease spread between individuals,} \begin{align}
\beta_{j \to i,t} = g(\bm{\beta},\bm{x}_{ijt}),
\end{align} where $g(\cdot)$ is a positive-valued function, $\bm{\beta}$ is a vector of unknown parameters, and $\bm{x}_{ijt}$ is a vector of covariates. For example, $\bm{x}_{ijt}$ could include the distance between individuals $i$ and $j$ \citep{almutiry2021continuous}, the age of individual $i$ or $j$ \citep{becker2017analysis}, or a temporal indicator for the implementation of some control measure \citep{lekone2006statistical}. 

Finally, we need to specify an initial state distribution for the Markov chain, that is, $P(S_{i0}=k)$ for $i=1,\dots,N$ and $k=1\text{(susceptible)},2\text{(infectious)},3\text{(removed)}$. This should be done based on knowledge of the outbreak, for example, based on who was likely the index case; see Section \ref{sec:noro_ex}.

\subsection{An autoregressive observation process for symptom-onset times} \label{sec:obs_model}

\begin{figure}[!t]
 	\centering
 	\includegraphics[width=.95\textwidth]{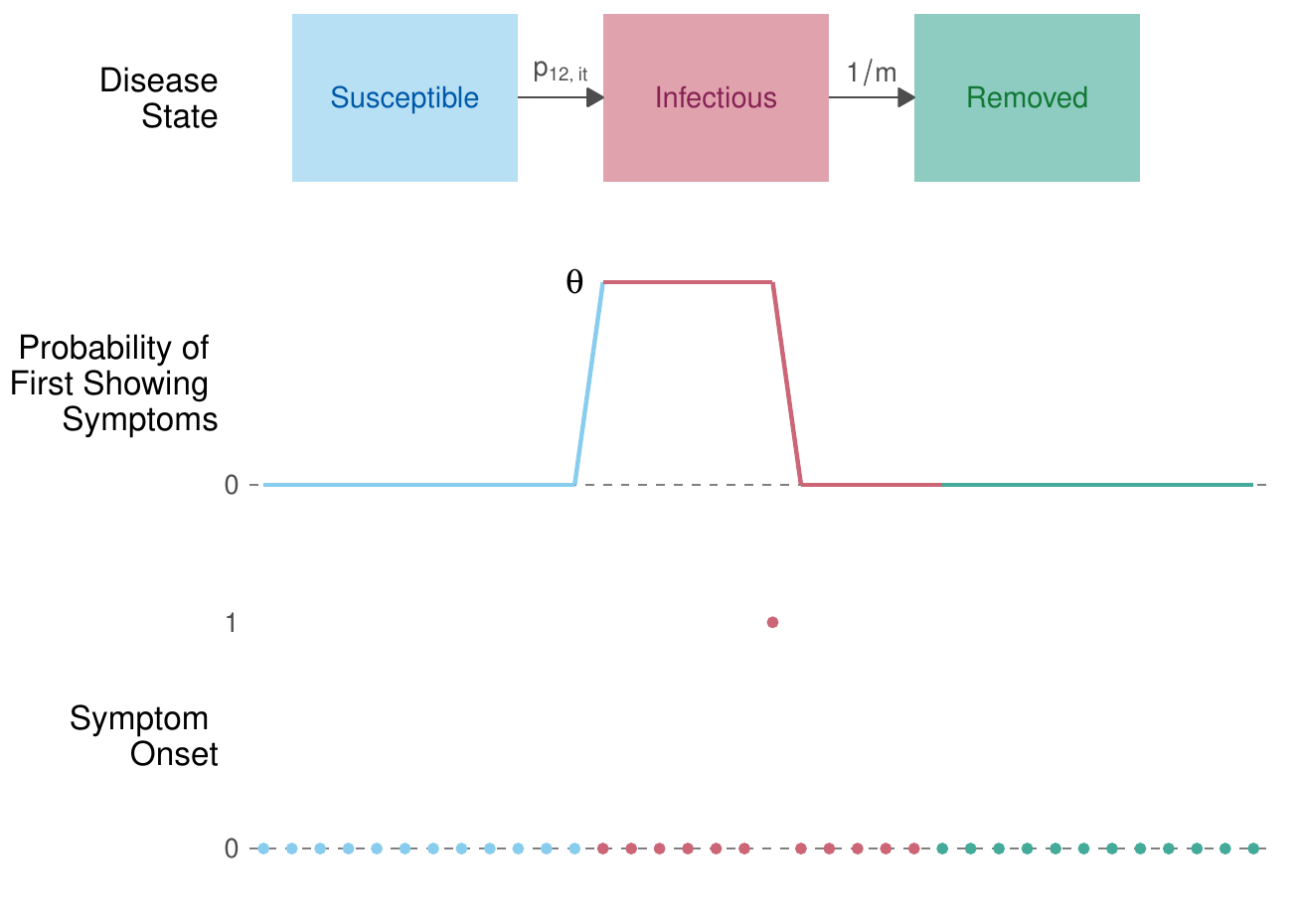}
	\caption{ Shows a diagram of the proposed hidden Markov individual-level model (HMM-ILM) for a hypothetical individual who showed symptoms during the study. The bottom graph shows the observations $y_{it}$, which denote symptom onset.\label{fig:model_diagram}} 
\end{figure}

A challenge with fitting the above transmission model to individual-level outbreak data is that we usually do not directly observe the transitions of $S_{it}$ {\color{black}(i.e., the infection and removal times)}. One solution is to specify an observation model that describes how what we do actually observe is related to the unknown underlying states \citep{touloupou_scalable_2020,pohlePrimerCoupledStateswitching2021}. {\color{black}In epidemiological studies of epidemics, we often only observe when symptoms first appeared in an individual \citep{neal_statistical_2004,kypraios_tutorial_2017,stockdale_modelling_2017,aristotelous_posterior_2023}.} Let $y_{it}=1$ if individual $i$ {\color{black}first showed symptoms of the disease during time $t$} and 0 otherwise. For {\color{black}individuals who showed symptoms,} $y_{it}$ will be equal to 0 until their {\color{black}symptom-onset} time, then it will be one, and then zero again until the end of the study; see the bottom graph of Figure \ref{fig:model_diagram}. For {\color{black}individuals who did not show symptoms}, $y_{it}=0$ for all $t$. {\color{black} Note that just because an individual did not show symptoms does not mean that they were not infected, as we explain at the end of the section.}

{\color{black} To derive the observation model, we will first assume that only infectious individuals show symptoms (i.e., that there are no false positives). This seems reasonable for our motivating examples, as we explain in Section \ref{sec:applications}. Suppose that after an individual becomes infected, they have a constant chance $\theta$ of first showing symptoms each time step, then we have} \begin{align} \label{eqn:GHMM_obs}
&y_{it} \mid S_{it},y_{i0},\dots, y_{i(t-1)}  \sim \begin{cases} 
        0, \hspace{.25cm} \text{if $S_{it} =1\text{ or 3}$ (susceptible or removed)}  \\[5pt]
     \text{Bern}(\theta\text{I}[y_{i0},\dots,y_{i(t-1)}=0]), \hspace{.25cm}  \text{if $S_{it} =2$ (infectious)},
   \end{cases}
\end{align} where $\text{I}[\cdot]$ is an indicator function. {\color{black}The indicator function in Equation (\ref{eqn:GHMM_obs}) will be 1 if the individual has not yet shown symptoms and 0 if they have already shown symptoms. Therefore, Equation (\ref{eqn:GHMM_obs}) states that the probability that an infectious individual first shows symptoms is $\theta$ if they have not yet shown symptoms and 0 if they have shown symptoms in a previous time step (an individual cannot show symptoms for the first time if they have already shown symptoms). In other words, by multiplying by the indicator function, we ensure that at most one symptom-onset time is observed per individual; see Figure \ref{fig:model_diagram}. Finally, we will assume that no individual showed symptoms before the study started, so that $y_{i0}=0$.} 

{\color{black} Note that Equation (\ref{eqn:GHMM_obs}) implicitly assumes that the incubation period (the time between infection and symptom onset) follows a geometric distribution with mean $1/\theta-1$ (we subtract 1 since an individual can show symptoms during the first time step of infectiousness). We give some ideas for relaxing this assumption to allow for more flexible incubation period distributions in Section \ref{sec:discussion}. There is generally good existing information about the average length of the incubation period \citep{caceres1998viral,brust2024}, which can be used to help form priors for $\theta$; see Section \ref{sec:noro_ex}.}

We will call the model defined by Equations (\ref{eqn:TMatrix})-(\ref{eqn:GHMM_obs}) the hidden Markov individual-level model (HMM-ILM). Note that the HMM-ILM, unlike most other epidemic models that employ DA methods \citep{o2019markov,bu_likelihood-based_2022}, does not assume that {\color{black}symptom onset corresponds to infection or removal}; see Figure \ref{fig:model_diagram}. {\color{black}Another advantage of the HMM-ILM is that, unlike the SINR approach in \cite{almutiry2021continuous}, it accounts for the possibility that an individual who did not show symptoms during the study was infected. In particular, following Equations (\ref{eqn:TMatrix}) and (\ref{eqn:GHMM_obs}), individuals can become infected and recover without showing symptoms, representing asymptomatic infections, which are common for many diseases \citep{wang2023global}.} Furthermore, individuals infected near the end of the study may not show symptoms before the study is over (e.g., an individual infected in the last time step would have only a $\theta$ chance of showing symptoms). {\color{black}If the outbreak is still ongoing at the end of the observation period, there will often be many recently infected individuals who are not yet showing symptoms \citep{jewell_bayesian_2009}. As we show in the next section, we treat the hidden states $S_{it}$ as unknown parameters of the model and place no constraints on them. Therefore, our inferential procedure will explore the parameter space consisting of all possible undetected infections (i.e., it will consider all possible infection patterns among individuals who did not show symptoms).

A final important advantage of the HMM-ILM is that, as we show in the next section, all hidden disease state indicators for an individual can be sampled directly from their full conditional distribution using the individual forward filtering backwards sampling (iFFBS) algorithm \citep{touloupou_scalable_2020}. More traditional DA methods have relied on complex Metropolis-Hastings proposals and reversible jump steps to sample unknown infection and removal times \citep{jewell_bayesian_2009}, which are difficult to implement and often do not mix well \citep{touloupou_scalable_2020,wang_bayesian_2025}.}

\section{Inferential Procedure} \label{sec:inf_proc}

{\color{black}The HMM-ILM is an autoregressive coupled hidden Markov model \citep{douwes-schultzZerostateCoupledMarkov2022a} (see Chapter 2 of \cite{douwes2024coupled} for a recent review). These models are similar to standard hidden Markov models except that the transition probabilities for one individual's chain can depend on the states of the other individuals; see Equation (\ref{eqn:prob_infect}). This between-chain dependency creates some challenges for inference, as we will now discuss.}

Let $\bm{v}=(\theta,m,\alpha,\bm{\beta})^T$ denote the vector of all model parameters, $\bm{S}_t=(S_{1t},\dots,S_{Nt})^T$ denote the vector of all disease state indicators at time $t$, $\bm{y}_{t}=(y_{1t},\dots,y_{Nt})^T$ denote the vector of all observations at time $t$, $\bm{S}=(\bm{S}_0,\dots,\bm{S}_T)^T$, and  $\bm{y}=(\bm{y}_1,\dots,\bm{y}_T)^T$. The joint distribution of $\bm{y}$ and $\bm{S}$ given $\bm{v}$ is given by
\begin{align} \label{eqn:joint_dist}
p\left(\bm{y},\bm{S}|\bm{v}\right) = \prod_{i=1}^{N}p\left(S_{i0}\right)\prod_{t=1}^{T} p\left(S_{it}|S_{i(t-1)},\bm{S_{(-i)(t-1)}},m,\alpha,
\bm{\beta}\right)p\left(y_{it}|y_{i0},\dots,y_{i(t-1)},S_{it},\theta\right).
\end{align} For {\color{black}individuals who showed symptoms}, we only observe the disease state indicator $S_{it}$ {\color{black}at their symptom onset time (then they must be infectious, as we assume no false positives).} {\color{black}For individuals who did not show symptoms,} we never observe $S_{it}$ {\color{black}(recall from the previous section that individuals who did not show symptoms may still have been infected)}. It is also not possible to marginalize $\bm{S}$ from (\ref{eqn:joint_dist}) to calculate $L(\bm{v})=p(\bm{y}|\bm{v})$, the marginal likelihood function, as doing so would require matrix multiplication with the $3^N$ by $3^N$ transition matrix of $\bm{S}_t$ \citep{douwes-schultzZerostateCoupledMarkov2022a}. Therefore, we treat the disease state indicators as unknown parameters of the model and sample $\bm{S}$ and $\bm{v}$ from their joint posterior distribution, which is proportional to
\begin{align} \label{eqn:postprop}
p(\bm{S},\bm{v}|\bm{y}) \propto p(\bm{y},\bm{S}|\bm{v})p(\bm{v}),
\end{align} where $p(\bm{v})$ is the prior distribution of $\bm{v}$. Regardless of the prior specification, the joint posterior distribution is not available in closed form. Therefore, we used a hybrid Gibbs sampling algorithm with some steps of the slice sampling {\color{black}or Metropolis-Hastings} algorithms to sample from it. We sampled $m$ and $\theta$ using {\color{black}either} univariate slice samplers \citep{neal2003slice} {\color{black} or adaptive random walk Metropolis steps \citep{valpineProgrammingModelsWriting2017}, depending on the mixing}. Since $(\alpha,\bm{\beta})^T$ showed high posterior correlations, we jointly sampled it using automated factor
slice sampling \citep{tibbits2014automated}. 

It is straightforward to sample the state indicators one at a time from their full conditional distributions, $p(S_{it}|\bm{S}\setminus\{S_{it}\},\bm{y},\bm{v})$ \citep{douwes-schultzZerostateCoupledMarkov2022a}. However, we found that this mixed so slowly in our two examples and simulation studies that it is not usable. Therefore, we instead jointly sampled each individual's disease state indicators, conditional on the disease state indicators for all other individuals, using the individual forward filtering backward sampling (iFFBS) algorithm \citep{touloupou_scalable_2020}. More specifically, the iFFBS algorithm samples the vector of all state indicators for individual $i$, $\bm{S}_{i(0:T)}=(S_{i0},\dots,S_{iT})^T$, from its full conditional distribution,  
\begin{align} \label{eqn:iFFBS}
p(\bm{S}_{i(0:T)}|\bm{S}_{(-i)(0:T)},\bm{y},\bm{v}) &= p(S_{iT}|\bm{S}_{(-i)(0:T)},\bm{y},\bm{v}) \prod_{t=0}^{T-1} p(S_{it}|S_{i(t+1)},\bm{S}_{(-i)(0:t+1)},\bm{y}_{i(0:t)},\bm{v}),
\end{align} where $\bm{S}_{(-i)(0:T)}$ is $\bm{S}$ with $\bm{S}_{i(0:T)}$ removed; see Supplemental Materials (SM) Section 1 for additional details. 

{\color{black}The} hybrid Gibbs sampling algorithm {\color{black}was implemented} using the R package NIMBLE \citep{valpineProgrammingModelsWriting2017}. NIMBLE comes with built-in slice and automated factor slice samplers. We implemented the iFFBS samplers using NIMBLE's custom sampler feature. All NIMBLE R code and data for our motivating examples are available on GitHub {\color{black}\url{https://github.com/Dirk-Douwes-Schultz/HMM_ILM_Code}}. In SM Section 2, we present a simulation study that shows our proposed Gibbs sampler can recover the true parameter values of an HMM-ILM, which is specified like in our TSWV example in Section \ref{sec:TSWV}.

Finally, we note that to run the iFFBS algorithm, we only require $S_{it}|\bm{S}_{(-i)(t-1)}$ to follow a first-order Markov chain and for $y_{it}$ to depend only on $S_{it}$ and, potentially, on past observations (see Section 2.2.1 of \cite{douwes2024coupled} for a list of exact conditions). Therefore, it is very easy to modify the HMM-ILM. For example, in Section \ref{sec:noro_ex}, for comparison purposes, we modify the observation process (\ref{eqn:GHMM_obs}) so that individuals are removed as soon as they {\color{black}show symptoms}  \citep{oneill_bayesian_1999}. 

\section{Applications} \label{sec:applications}
To fit all models discussed in this section, we ran our hybrid Gibbs sampling algorithm for 200,000 iterations on three chains with an initial burn-in of 50,000 iterations. All chains were started from random values in the parameter space to avoid convergence to local modes. Convergence
was checked using the Gelman–Rubin statistic (all estimated parameters<1.05), the minimum effective sample size (>1000), and by visually examining the traceplots \citep{plummerCODAConvergenceDiagnosis2006}. Unless indicated otherwise, {\color{black}no convergence issues were apparent for} any model discussed in this section.

\subsection{Norovirus outbreak in a South Carolina hospital}
\label{sec:noro_ex}

\begin{figure}[!t]
 	\centering
 	\includegraphics[width=.7\textwidth]{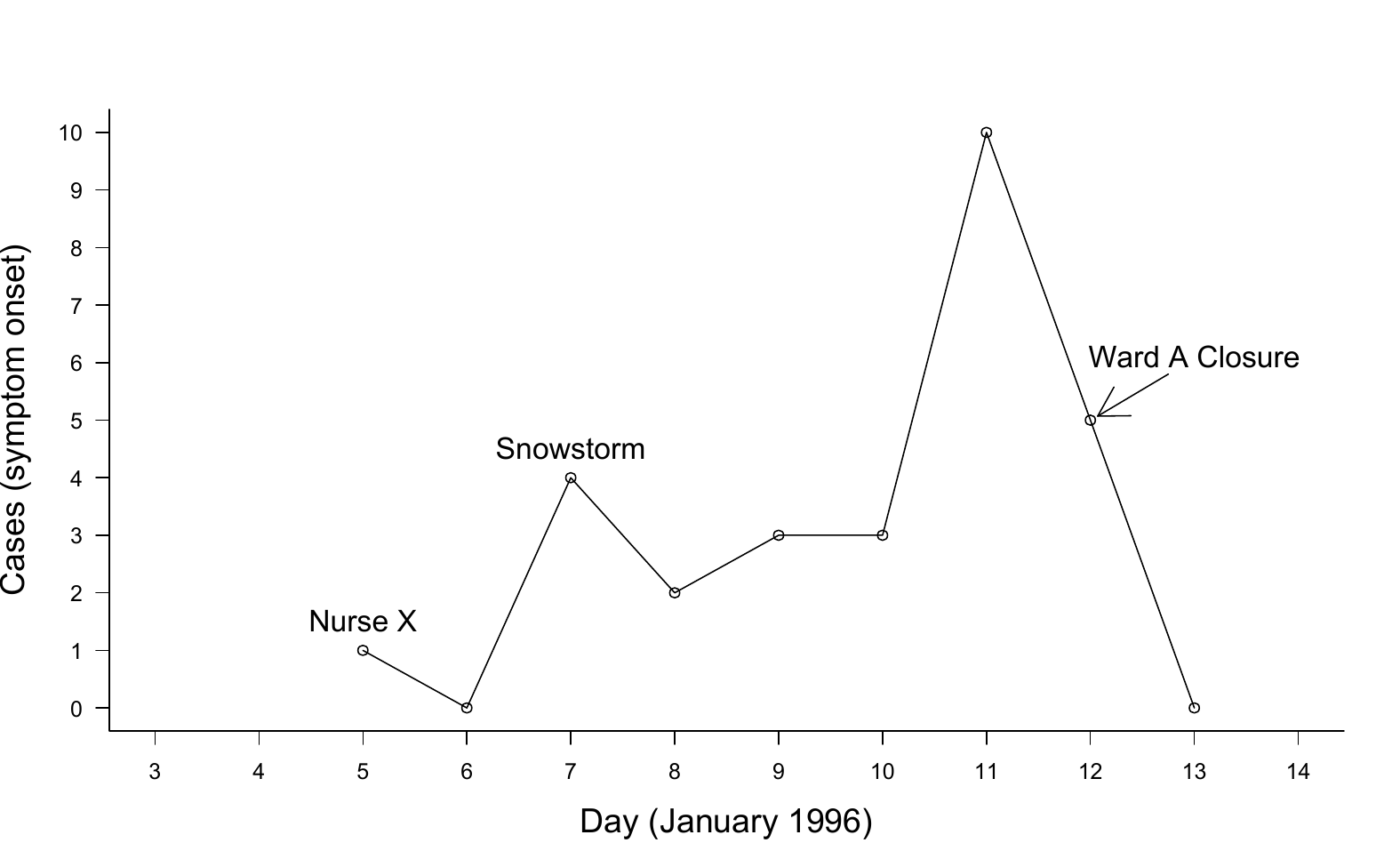}
	\caption{ Cases in nurses by day of symptom onset. Norovirus outbreak in a South Carolina hospital. \label{fig:noro_epi_curve}} 
\end{figure}

For our first example, we analyze a norovirus outbreak among 89 nurses in a South Carolina hospital in January 1996 \citep{caceres1998viral}. The observation period for the outbreak investigation extended from the {\color{black}5th} to the 13th (t=1,\dots,T=9). The data consist of 28 {\color{black}symptom-onset times}, which correspond to the day the nurse first reported showing symptoms in a survey administered by investigators. Figure \ref{fig:noro_epi_curve} shows the epidemic curve, that is, the number of new cases over time. The first case was Nurse X on the fifth. Nurse X was thought to be the nurse who introduced the infection into the ward, as many of the early cases were traced back to her. The ward was closed on the 12th, and after that, no new cases were reported. 

 \cite{kypraios_tutorial_2017} previously analyzed this data using a homogeneously mixing SIR model. However, they assumed that all the removal times in the interval $[1,T]$ had been observed and were equal to the symptom-onset times, which is a common assumption following the influential work of \cite{oneill_bayesian_1999}. That is, they assumed that Nurse X was removed on the fifth, that the four nurses detected on the seventh were removed that day, and so on. However, a snowstorm on the seventh forced the nurses to stay overnight at the hospital, so they could not have been removed that day. Also, according to the investigation report \citep{caceres1998viral}, many nurses continued to work after developing symptoms due to staff shortages. Additionally, since \cite{kypraios_tutorial_2017} assumed they had observed all removal times in the interval $[1,T]$, they implicitly assumed that none of the 61 nurses {\color{black}who did not show symptoms} had been removed during the observation period. This is an often overlooked issue with the framework of \cite{oneill_bayesian_1999}. Norovirus can frequently have mild or no symptoms \citep{wang2023global}, meaning it is not unreasonable to assume that at least some of the nurses were infected and recovered without {\color{black}showing symptoms}.

In contrast, the HMM-ILM allows individuals to be removed even several days after {\color{black}symptom onset}, as illustrated in Figure \ref{fig:model_diagram}. Additionally, the HMM-ILM allows for {\color{black}nurses to become infected and recover without showing symptoms}. Therefore, it may be more suitable for this data. As we have no individual-level information, we used a simple homogeneous mixing model to describe the effects of disease spread between the nurses,
\begin{align} \label{eqn:hospital_infect}
\beta_{j \to i,t} &= \beta (1-{\color{black}\text{W}_{t-1}}),
\end{align} where {\color{black}$\text{W}_t$} is an indicator for whether the ward was closed that day. That is, we are assuming that if the ward is closed, the nurses will not mix with each other. Given that there is no obvious neighbourhood structure, we let $NE(i)=\{j : j \neq i\}$. {\color{black}Recall that the HMM-ILM assumes that only those with norovirus showed symptoms. The symptoms in this case are combined vomiting and diarrhea, which are shared with food poisoning. However, the investigators tested a subset of nurses for many common causes of food poisoning, and all tested negative \citep{caceres1998viral}.} Finally, for the initial state distributions, we assumed that Nurse X had a 95\% chance of being initially infected, as they likely introduced the infection into the ward \citep{caceres1998viral}. We assumed that all other nurses had a low 1\% chance of being initially infected.

We first tried to fit the HMM-ILM specified above using wide priors for all parameters. However, with only 28 {\color{black}symptom-onset} times, the posteriors were very uninformative, and the model had trouble converging. Therefore, we decided to use {\color{black}existing epidemiological knowledge} to specify informative priors for $\theta$ and $m$. Norovirus has a short incubation period of 1-2 days \citep{caceres1998viral}, so we placed a $\text{Beta}(40,40(6/4))$ prior on $\theta$ {\color{black}(see Section \ref{sec:obs_model})}. For $m$, an individual should be infectious with norovirus for an average of 12 days without any intervention \citep{lappe2023predicting}. However, it is unlikely that infectious nurses would have worked at the hospital for this entire duration. For example, the investigation report talks about a nurse who first showed symptoms on the seventh, stayed overnight due to the snowstorm, worked the next day, and was not removed until the ninth. Therefore, we assume $m$ is likely somewhere between 1.5 (average time to symptom onset) and 12 days, and we placed a $\text{Gamma}(2,2/5.75)$ prior on $m-1$. Finally, we placed wide $\text{Unif}(0,1)$ priors on $\beta$ and $\alpha$, from Equations (\ref{eqn:prob_infect}) and (\ref{eqn:hospital_infect}). 

\begin{figure}[!t]
 	\centering
 	\includegraphics[width=.75\textwidth]{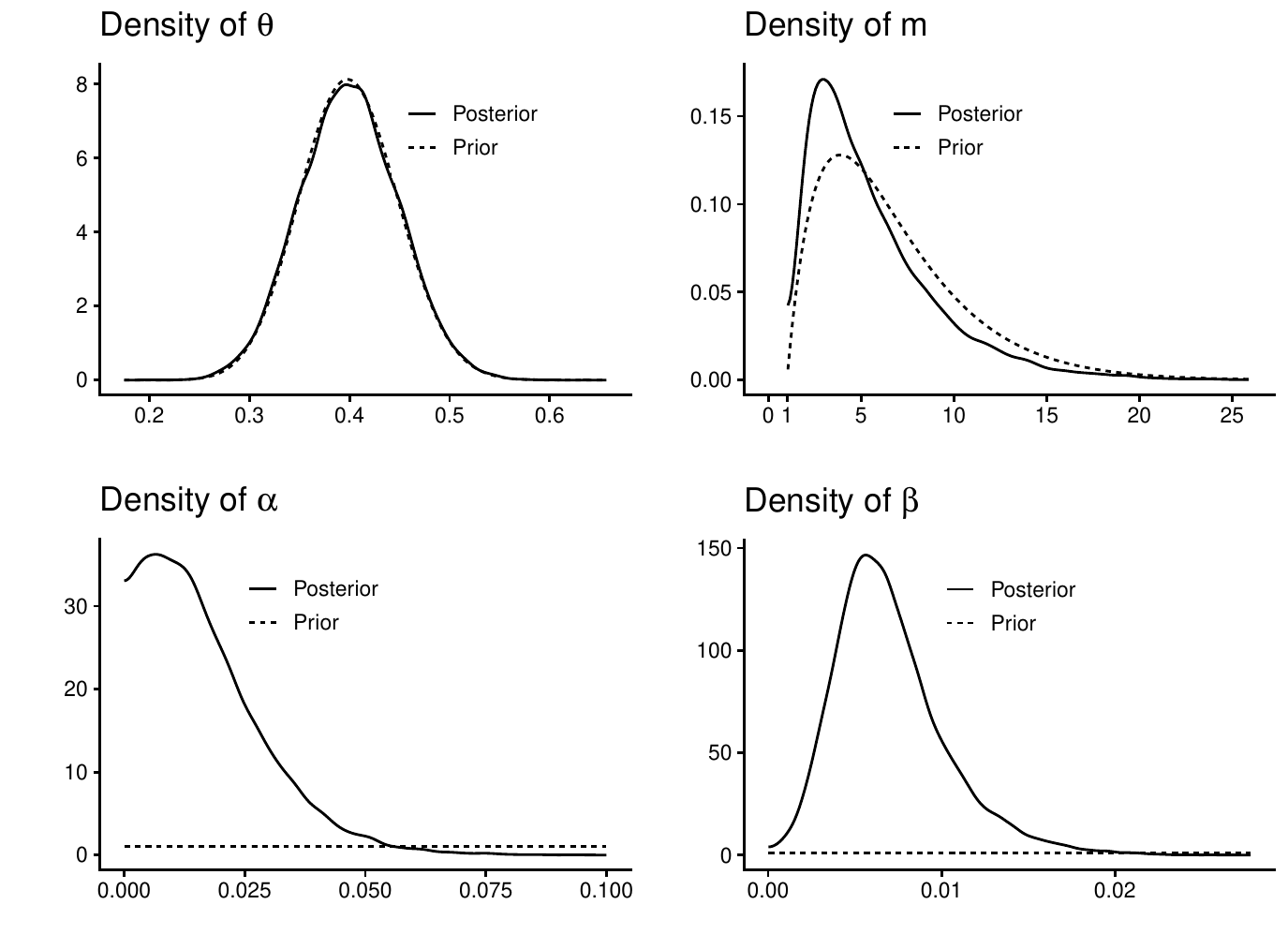}
	\caption{Posterior and prior distributions of $\theta$, $m$, $\alpha$, and $\beta$ from fitting the HMM-ILM to the norovirus outbreak data.\label{fig:prior_post_noro}} 
\end{figure}

Figure \ref{fig:prior_post_noro} shows the prior and posterior distributions of  $\theta$, $m$, $\alpha$, and $\beta$. The prior and posterior distributions of $m$ and, especially, $\theta$ are very similar, indicating that there is not a lot of information about these parameters in the data. In contrast, the data is much more informative about $\alpha$ and $\beta$. From the posteriors, we estimate that there was a daily background infection risk of 1.4\% (0.08\%, 4.8\%) (posterior median and 95\% credible interval), and that the probability of an infectious contact between a susceptible and infectious nurse was 0.65\% (0.2\%, 1.5\%) per day. This implies that an infectious nurse would produce an average of 2.61 (1, 7.04) secondary infections before recovering in a completely susceptible population, which is known as the basic reproduction number $R_0=(N-1)\beta m$ \citep{vynnycky2010introduction}. The background infection risk could be attributed to the 10 patients who were infected on the ward during the observation period and are not typically included in the model \citep{britton_bayesian_2002,kypraios_tutorial_2017}. It may also represent infections that are not well explained by the assumptions of the model, such as homogeneous mixing.

We can estimate {\color{black}how many of the 61 symptomless nurses} were removed during the observation period by looking at the posterior median and 95\% credible interval of $\sum_{i=29}^{89}I[S_{iT}=3]$, which is 9 (1, 31) (we let nurses 29-89 be the {\color{black}symptomless} ones). Therefore, it is likely that at least a few nurses were infected and recovered without {\color{black}showing symptoms}. This casts some doubt on the assumption made by \cite{kypraios_tutorial_2017} that none of these nurses were removed. However, the credible interval, while not including 0, is still very wide with only 28 {\color{black}symptom-onset} times. Finally, we can better understand how the outbreak evolved over time by examining the posterior probability that a nurse was in each epidemiological state during each day. We do this for one of the 4 nurses {\color{black}with symptom onset} on the seventh in SM Figure 5. From the figure, this nurse was likely infected on the fifth or sixth and {\color{black} not} removed {\color{black}until at least} the ninth or tenth, which corresponds to the description of one of the nurses {\color{black} with onset }on the seventh given in the investigation report (as discussed above). This again casts some doubt on {\color{black}the assumption made in \cite{kypraios_tutorial_2017} that} this nurse was removed on the seventh and would not have been able to infect anyone else on subsequent days. Examining plots of the other nurses showed a similar pattern of {\color{black}at least} 2-4 days between {\color{black}symptom onset} and removal.

\begin{figure}[!t]
 	\centering
 	\includegraphics[width=\textwidth]{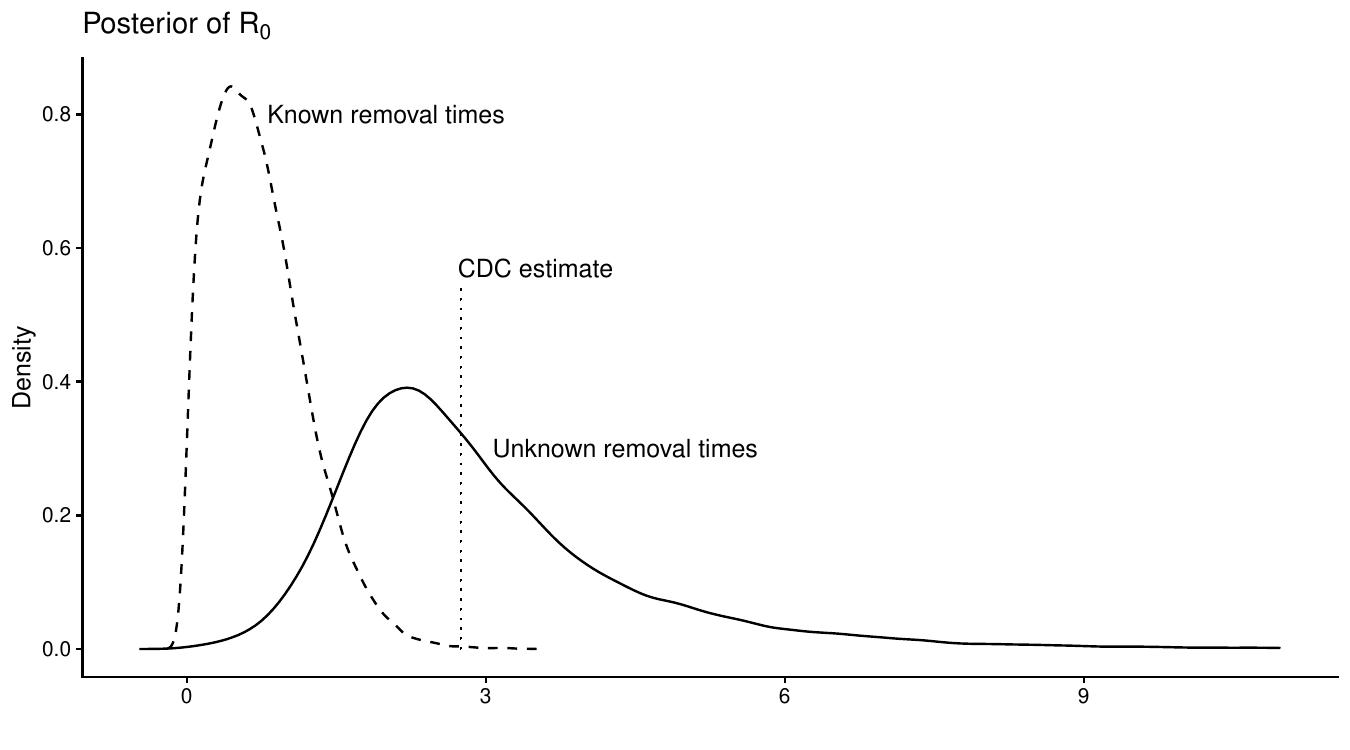}
	\caption{ Shows the posterior distribution of $R_0$ for a model that assumes unknown removal times, the HMM-ILM, and a model that assumes known removal times, that is, that the removal times are equal to the symptom-onset times as in \cite{kypraios_tutorial_2017}. The vertical dotted line is drawn at the median $R_0$ estimate from \cite{steele2020characterizing}.\label{fig:R0_post_compare}} 
\end{figure}

We can incorporate known removal times by modifying the observation process (\ref{eqn:GHMM_obs}),
\begin{align} \label{eqn:GHMM_obs_KRT}
&y_{it} \mid S_{it},y_{i0},\dots, y_{i(t-1)}  \sim \begin{cases} 
        0, \hspace{.25cm} \text{if $S_{it} =1\text{ or 2}$ (susceptible or infectious)}  \\[5pt]
    {\color{black} \text{Bern}(\text{I}[y_{i0},\dots,y_{i(t-1)}=0])}, \hspace{.25cm}  \text{if $S_{it} =3$ (removed)}.
   \end{cases}
\end{align} Note (\ref{eqn:GHMM_obs_KRT}) assumes we have observed the removal time of every nurse removed in the interval $[1,T]$, and that the removal times are equal to the symptom onset times, as in \cite{kypraios_tutorial_2017} and \cite{oneill_bayesian_1999} (see above for an indepth discussion of these assumptions). Figure \ref{fig:R0_post_compare} compares the posterior distribution of $R_0$ between the HMM-ILM described in the previous paragraphs and an HMM-ILM where (\ref{eqn:GHMM_obs}) is replaced by (\ref{eqn:GHMM_obs_KRT}), which we will refer to as the known removal times (KRT) model. Assuming known removal times, a nurse's removal time is equal to their symptom onset time, meaning we can interpret $m$ as the average duration {\color{black}of the incubation period}. Therefore, we placed a $\text{Unif}(1,3)$ prior on $m$ for the KRT model based on the {\color{black}existing epidemiological knowledge} discussed above. We kept the other model specifications the same.

From Figure \ref{fig:R0_post_compare}, assuming known removal times results in a much smaller and more concentrated posterior for $R_0$. A CDC study of 7,094 norovirus outbreaks found a median $R_0$ estimate of 2.75 (interquartile range 2.38-3.65) \citep{steele2020characterizing}, which is not well represented by the posterior from the KRT model. The posterior median and 95\% credible interval of the intercept $\alpha$ was 3.28\% (0.53\%, 6.64\%) for the KRT model, while, recall, it was 1.4\% (0.08\%, 4.8\%) for our HMM-ILM. A large intercept like this may indicate model mispecification, since it implies that many of the infections cannot be explained by mixing between the nurses. The KRT model attributes a significant portion of the infection risk to background reservoirs, while producing an unrealistically low estimate of the risk from between-nurse mixing, as represented by $R_0$. 

The two models {\color{black}can also be compared} using the widely applicable information criterion (WAIC) \citep{gelman2014understanding,douwes2025three}; see SM Section 3. The model with the smallest WAIC is considered to have the best fit, and as a rule of thumb, a difference of 5 or more in the WAIC is considered significant \citep{reich2019bayesian}. {\color{black}The} HMM-ILM had a WAIC of 221, while the KRT model had a WAIC of 222.36. Since the difference in WAIC is less than 5, this comparison is inconclusive. In SM Section 4.2, {\color{black}we carry out a sensitivity analysis in which} we fit several alternative models, including those without an intercept or assuming the outbreak had ceased, and we were still unable to find any significant differences in the WAIC. This suggests that the sample size of only 28 {\color{black}symptom-onset} times may be too small to distinguish between different epidemic models using model comparison criteria. This is despite the fact that important model results can be very sensitive to assumptions about the observation process, as shown in Figure \ref{fig:R0_post_compare}.  Therefore, caution should be exercised when analyzing smaller outbreaks. If we cannot compare models using comparison criteria, we have to rely more heavily on our a priori assumptions about the outbreak, which can have a large impact on the results. For this example, even though the WAIC is inconclusive, we would argue our HMM-ILM more closely matches the description of the outbreak given in \cite{caceres1998viral} compared with the KRT model{\color{black},} and {\color{black}it} also produces more realistic estimates, as discussed above. Note that small sample sizes are common when fitting ILMs to incomplete epidemic data \citep{oneill_bayesian_1999,britton_bayesian_2002,stockdale_modelling_2017,stockdale_pair-based_2021} (all with 40 or fewer {\color{black}symptom onset} times).

As we show in the next example, when the sample size is larger, 327 {\color{black}symptom-onset} times, we often do obtain significant differences in the WAIC. This is true even for models that have similar specifications (like a linear versus a non-linear distance kernel). Additionally, we can use much wider priors for the model parameters and still obtain informative posteriors. Therefore, this example needs to be viewed in the context of its small sample size.

\subsection{Experiment on the spread of tomato spotted wilt virus in pepper plants} \label{sec:TSWV}

\begin{figure}[!t]
 	\centering
 	\includegraphics[width=\textwidth]{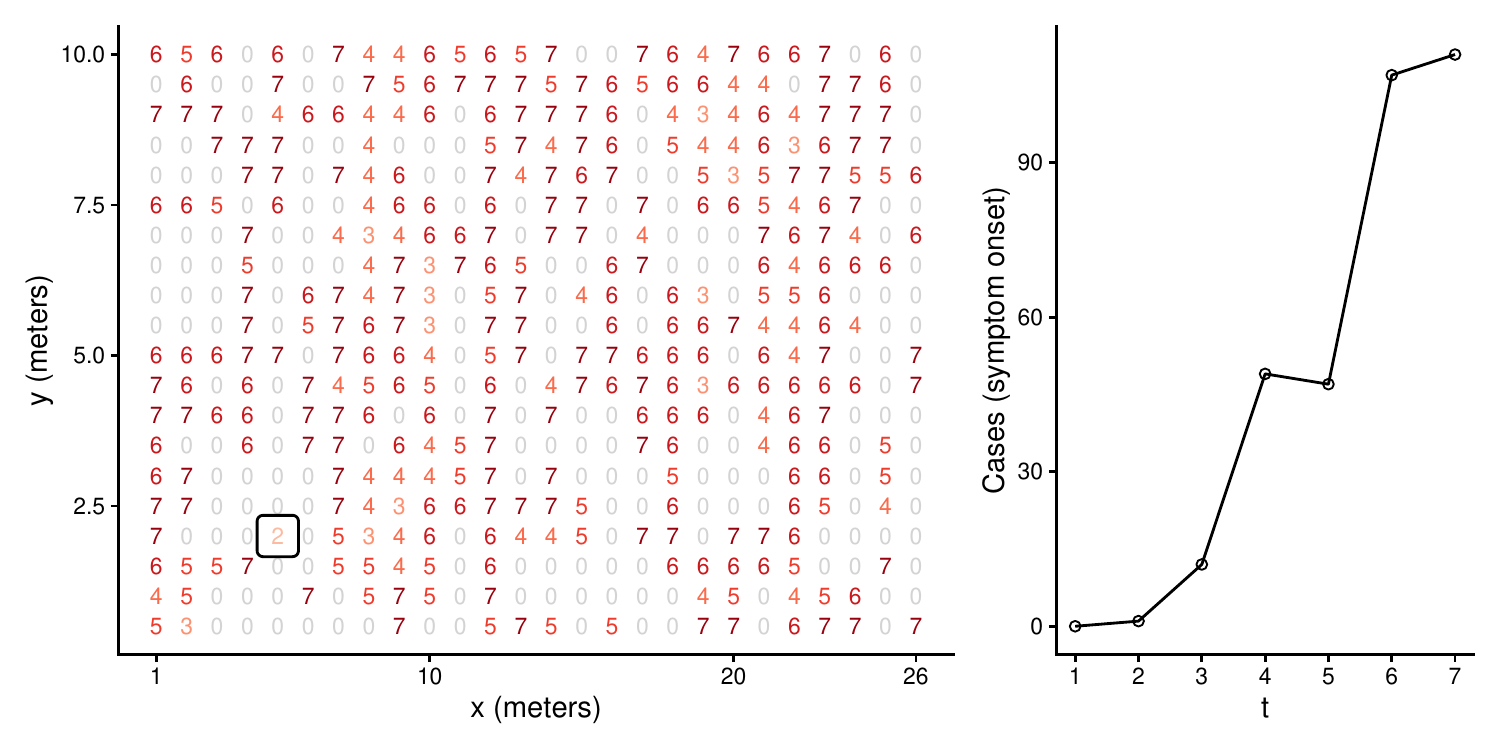}
	\caption{ The left plot shows the symptom-onset time of each plant that showed symptoms in the experiment from \cite{hughes1997validating}. A box is drawn around the first plant that showed symptoms (at time $t=2$). A 0 corresponds to the plant never showing symptoms during the experiment. The right plot shows the epidemic curve.\label{fig:TSWV_detection_times} }
\end{figure}

For our second example, we analyze an experiment on the spread of tomato spotted wilt virus (TSWV) in pepper plants, which was described in \cite{hughes1997validating}. In this experiment, they had $N=520$ pepper plants planted in 26 rows of 20 plants each. Plants were checked for symptoms of TSWV every 2 weeks for 14 weeks ($t=1,\dots,T=7$). The data consist of {\color{black}the symptom-onset time of each of the 327 plants that showed symptoms during the experiment}. The {\color{black}symptom-onset} times are plotted on a spatial grid in Figure \ref{fig:TSWV_detection_times}. 

Many previous attempts to analyze this data have assumed that the infection times are known and equal to the symptom-onset times \citep{pokharel2014supervised,pokharel2016gaussian,v_individual-level_2020,bilodeau2024stochastic}. However, it takes 2-4 weeks for symptoms of TSWV to appear in a plant \citep{brust2024}, meaning that plants were likely infected {\color{black}in a previous biweek before they first showed symptoms}. \cite{almutiry2021continuous} analyzed this data using {\color{black}an} SINR model (see the introduction) and estimated the unknown infection and removal times of the 327 plants {\color{black}that showed symptoms}. However, they assumed that all 193 plants {\color{black}that did not show symptoms during the experiment escaped infection}. Some of the plants {\color{black}that did not show symptoms} could have been infected near the end of the experiment and would have perhaps shown symptoms if the observation period had been extended (it takes 2-4 weeks for symptoms to develop). Indeed, the epidemic curve, plotted in Figure \ref{fig:TSWV_detection_times}, is at its peak, indicating that we would likely see more cases if the experiment had gone on longer. Also, some {\color{black}symptomless} plants could have had asymptomatic infections \citep{mullis2009tospoviruses}.

In contrast, the HMM-ILM {\color{black}accounts for the random delay between infection and symptom onset}. Furthermore, the HMM-ILM will account for the possibility that {\color{black}a plant that did not show symptoms was infected.} Therefore, an HMM-ILM may be more suitable for this data. {\color{black}The HMM-ILM assumes that only plants infected with TSWV showed symptoms. Given that it is a planned experiment and the symptoms on the pepper
are fairly distinctive \citep{nsw_tomato_spotted_2024}, this seems like a reasonable assumption.} 

TSWV is transmitted from plant to plant by small insects called thrips \citep{mullis2009tospoviruses}. It seems reasonable to assume that the thrips are more likely to move to closer plants compared to those farther away. Therefore, we used a second-order Taylor series approximation of the power-law distance kernel \citep{deardon_inference_2010} to describe the effect of disease spread between plants (see Equation (\ref{eqn:prob_infect})),
\begin{align} \label{eqn:plTaylor}
\begin{split}
\beta_{j\to i,t}=\beta_{j \to i}&=\beta_0\beta_1^{-d_{ij}} \\
& \approx \beta_0d_{ij}^{-a}\left(1-\ln(d_{ij})(\beta_1-a)+.5\ln(d_{ij})^2(\beta_1-a)^2\right),
\end{split}
\end{align} where $d_{ij}$ is the distance (in meters) between plants $i$ and $j$. In (\ref{eqn:plTaylor}), $\beta_0>0$ represents the effect of disease spread from plants one meter away, while $\beta_1>0$ represents how fast the effect of disease spread decays with the distance between the two plants. We used a Taylor series approximation as it allows us to factor $\beta_0$ and $\beta_1$ from the sums $\sum_{j \in NE(i)}\beta_{j \to i,t}I[S_{j(t-1)}=2]$ in Equation (\ref{eqn:prob_infect}). This means we do not need to update these sums when updating $\beta_0$ and $\beta_1$ in our MCMC algorithm, which saves a lot of computational time. Note that the Taylor series approximation is always positive. We set $a=1.35$, which is a guess for $\beta_1$, from \cite{v_individual-level_2020}. In SM Section 5.1, we investigate some alternative distance kernels, including linear kernels, $\beta_{j\to i}=\beta_0+\beta_1d_{ij}$; quadratic kernels, $\beta_{j \to i}=\beta_0+\beta_1d_{ij}+\beta_2d_{ij}^2$ (with positivity constraints); and spline-based kernels. We found the HMM-ILM with {\color{black}Equation} (\ref{eqn:plTaylor}) had the lowest WAIC by a significant amount, {\color{black}so} we use {\color{black}this expansion} throughout this section. We also compared the Taylor approximation with the exact power-law and found they had the same WAIC, meaning the {\color{black}Taylor series} approximation does not {\color{black}appear to} negatively affect the fit.

We used a wide prior for $\theta \sim \text{Unif}(0,1)$, $m\sim \text{Unif(1,20)}$, $\alpha \sim \text{Unif}(0,1)$, $\beta_0 \sim \text{Unif}(0,1)$, and $\beta_1 \sim \text{Unif}(0,20)$. As it is not known how the disease was introduced in the experiment, we assumed each plant had a low 1\% chance of being initially infected. Finally, concerning the neighbourhood set $NE(i)$, there is likely a maximum distance we would expect the thrips to be able to travel within 2 weeks. Therefore, we started with a second-order {\color{black}queen} neighborhood structure (all plants at most 2 plants away{\color{black}, including on the diagonal}) {\color{black}\citep{nguyen2025assessing}} and increased the neighbourhood order by one until the WAIC decreased by less than 5. The second-order neighbourhood model had a WAIC of 1602.68, the third-order model had a WAIC of 1591.76, and the fourth-order model had a WAIC of 1587.32. Therefore, we let $NE(i)$ include all plants at most three plants away from plant $i$ {\color{black}(see SM Figure 7)}.

\begin{figure}[!t]
 	\centering
 	\includegraphics[width=\textwidth]{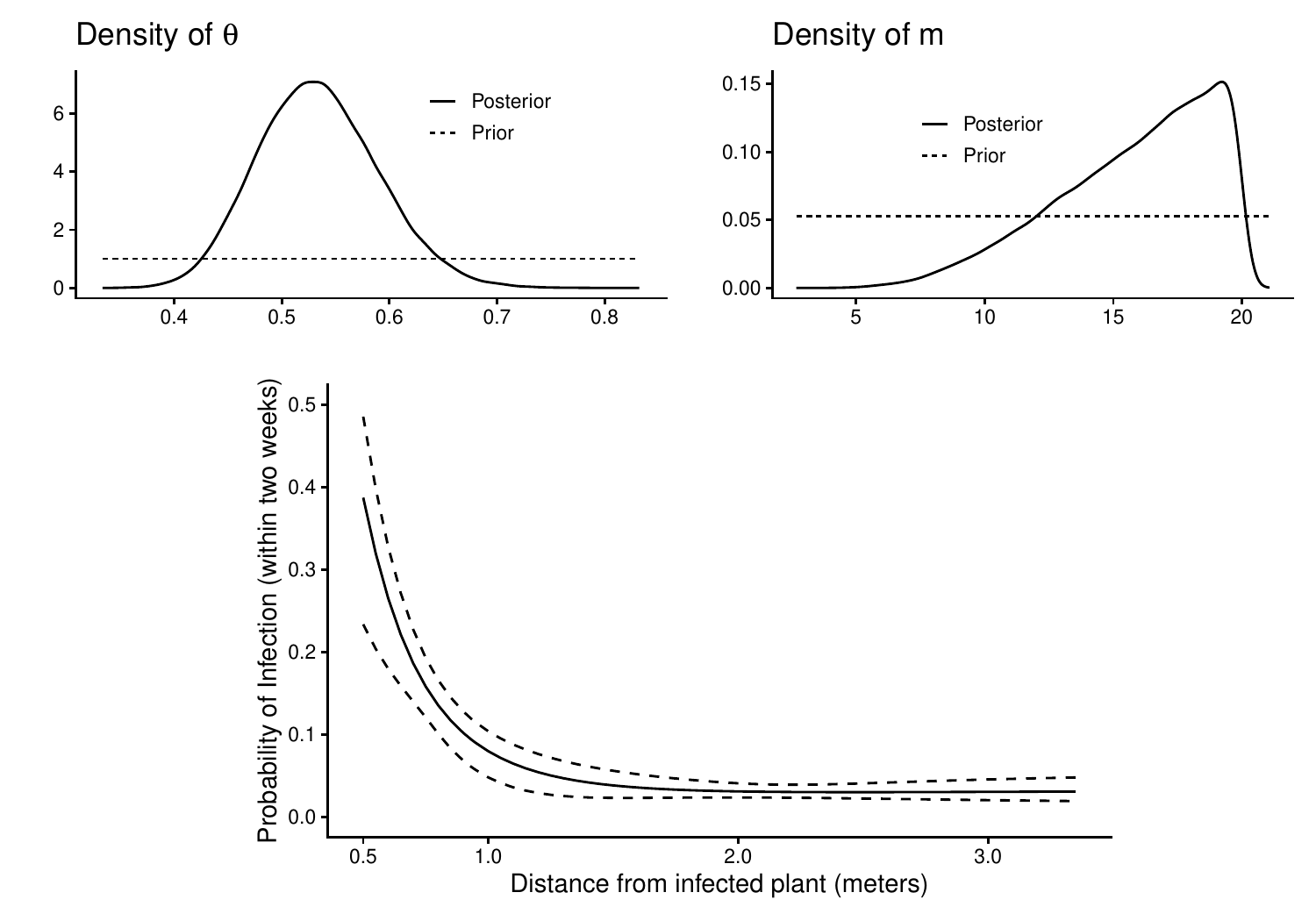}
	\caption{ The top graphs show the posterior and prior distributions of $\theta$ and $m$ from fitting our HMM-ILM to the TSWV experiment. The bottom graph shows the probability that a susceptible plant is infected (within two weeks), given a single infectious plant, as a function of the distance between the two plants (posterior median and 95\% credible interval).\label{fig:TSWV_main} }
\end{figure}

The top graphs of Figure \ref{fig:TSWV_main} show the prior and posterior distributions of $\theta$ and $m$. The posterior median and 95\% credible interval of $\theta$ is 0.53 (0.43, 0.65), meaning we estimate that it took an infected plant 1.77 (1.08, 2.65) weeks to {\color{black}show symptoms} {\color{black}on average}. This corresponds with the known incubation period \citep{brust2024}. Although the posterior of $m$ is wide, there is a 95\% chance that the average duration of the infectious period is at least 20 weeks. Plants are infected with TSWV until they die \citep{nsw_tomato_spotted_2024}, so a long infectious period like this seems plausible. In our simulation study in SM Section 2, we found that when the true value of $m$ is greater than $T$, the posterior is often wide {\color{black}as seen} in Figure \ref{fig:TSWV_main}. In contrast, when the true value of $m$ is less than $T$, the posterior is usually much more concentrated around the true value. This suggests that when the average duration of the infectious period is greater than the length of the observation period, it can be challenging to estimate $m$.

The bottom graph of Figure \ref{fig:TSWV_main} shows the probability that a susceptible plant is infected (within two weeks), given a single infectious plant, versus the distance between the two plants. For example, if a susceptible plant is half a meter from an infectious plant (the minimum distance in the experiment), it would have a 39\% (23\%, 49\%) chance of being infected within two weeks. If the plant were 2 meters away, it would have only a 3\% (2\%, 4\%) chance of being infected. Therefore, we can see from the figure that the probability of infection decays very rapidly with the distance from an infected plant. In particular, susceptible plants more than 1.5 meters from an infected plant are unlikely to become infected within 2 weeks, so the range of dispersal is fairly short. Note that the exact equation plotted, from Equations (\ref{eqn:prob_infect})+(\ref{eqn:plTaylor}), is $1-\exp\left(-\alpha-\beta_0d_{ij}^{-a}\left(1-\ln(d_{ij})(\beta_1-a)+.5\ln(d_{ij})^2(\beta_1-a)^2\right)\right)$ versus $d_{ij}$.

\begin{figure}[!t]
 	\centering
 	\includegraphics[width=\textwidth]{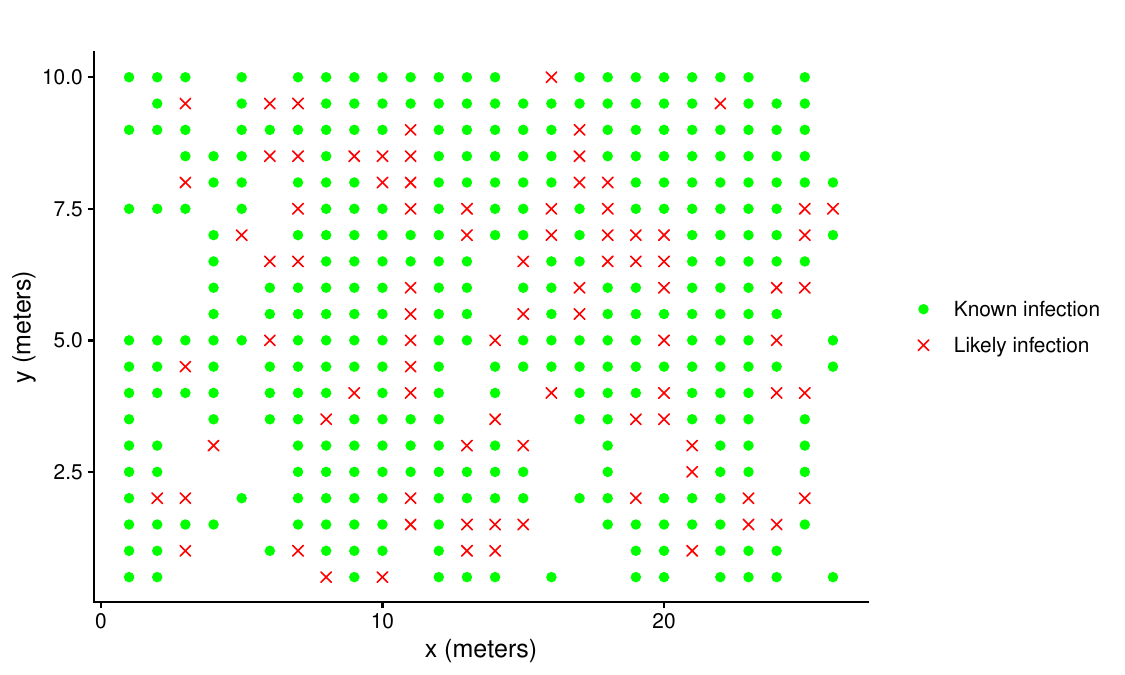}
	\caption{ The circles represent plants that {\color{black}showed symptoms} during the experiment {\color{black}and thus are known to have been infected}. The crosses represent plants {\color{black}that did not show symptoms but} whose posterior probability of having been infected during the experiment, $P(S_{iT}=\text{2 or 3}|\bm{y})$, is greater than .5. \label{fig:TSWV_likely_infections} }
\end{figure}

The posterior probability that a plant $i$ was infected during the experiment is given by $P(S_{iT}=\text{2 or 3}|\bm{y}) \approx \frac{1}{Q-M}\sum_{q=M+1}^{Q}I[S_{iT}^{[q]}=\text{2 or 3}]$, where the superscript $[q]$ denotes a draw from the posterior distribution of the variable, $Q$ is the total number of MCMC samples, and $M$ is the size of the burn-in sample. {\color{black}If a plant showed symptoms, then $P(S_{iT}=\text{2 or 3}|\bm{y})=1$, that is, we know it was infected. If a plant did not show symptoms, recall that there is still a chance it was infected during the experiment, which is captured by $P(S_{iT}=\text{2 or 3}|\bm{y})$. Therefore, we can identify plants that were likely infected but never showed symptoms (i.e., undetected infections) by finding plants for which $y_{it}=0$ for all $t$ and $P(S_{iT}=\text{2 or 3}|\bm{y})>.5$}. We plot the location of such plants in Figure \ref{fig:TSWV_likely_infections} (denoted by crosses). From the figure, we can see that many plants {\color{black}that did not show symptoms but were} in close proximity to symptomatic plants were likely infected during the experiment. These plants may have had asymptomatic infections or were infected late in the experiment and were not yet showing symptoms. We can estimate the total number of plants {\color{black}that were infected but never} showed symptoms during the experiment by looking at the posterior median and mean of $\sum_{i\,:\,y_{it}=0 \,\forall \,t} I[S_{iT}=\text{2 or 3}]$. {\color{black}We find this to be} 94 (58, 133). Therefore, even on the low end of our estimation, we would expect {\color{black}58/193 of the plants that never showed symptoms} to have been infected during the experiment. This {\color{black}suggests that the analysis from} \cite{almutiry2021continuous}, who assumed that none of these plants were infected, {\color{black} was based on over-simplified assumptions}.

For comparison purposes, we can incorporate the assumption that {\color{black}all plants that did not show symptoms escaped infection}, as in \cite{almutiry2021continuous}, by fixing $S_{it}=1$ (susceptible) for all plants $i$ {\color{black}that did not show symptoms} and $t=0,\dots,7$. We can incorporate known infection times, like in \cite{bilodeau2024stochastic} or \cite{v_individual-level_2020}, by modifying the observation process (\ref{eqn:GHMM_obs}), 
\begin{align} \label{eqn:GHMM_obs_KIT}
&y_{it} \mid S_{it},y_{i0},\dots, y_{i(t-1)}  \sim \begin{cases} 
        0, \hspace{.25cm} \text{if $S_{it} =1\text{ or 3}$ (susceptible or removed)}  \\[5pt]
    {\color{black} \text{Bern}(\text{I}[y_{i0},\dots,y_{i(t-1)}=0])}, \hspace{.25cm}  \text{if $S_{it} =2$ (infectious)}.
   \end{cases}
\end{align} Note Equation (\ref{eqn:GHMM_obs_KIT}) assumes we have observed the infection time of every plant infected in the interval $[1,T]$, and that the infection times are equal to the {\color{black}symptom-onset times}. Finally, we also want to consider the {\color{black}independent Bernoulli observation model} of \cite{touloupou_scalable_2020}, since it could be naively applied in this application {\color{black}as the observations $y_{it}$ are binary},
\begin{align} \label{eqn:GHMM_obs_cont}
&y_{it} \mid S_{it}  \sim \begin{cases} 
        0, \hspace{.25cm} \text{if $S_{it} =1\text{ or 3}$ (susceptible or removed)}  \\[5pt]
    {\color{black} \text{Bern}(\theta)}, \hspace{.25cm}  \text{if $S_{it} =2$ (infectious)}. 
   \end{cases}
\end{align}  {\color{black}To understand the problem with this model, suppose that individual $i$ first shows symptoms during time $t^{*}$, so that $y_{it^*}=1$, and that this individual is still infectious at time $t^*+1$, so that $S_{i(t^*+1)}=2$. Then, under Equation (\ref{eqn:GHMM_obs_cont}), the probability that $y_{i(t^*+1)}=1$ is equal to $\theta$. Therefore, an immediate issue with (\ref{eqn:GHMM_obs_cont}) is that it can generate multiple symptom-onset times for an individual, provided $m>1$ and $\theta>0$.}

\begin{figure}[!t]
 	\centering
 	\includegraphics[width=\textwidth]{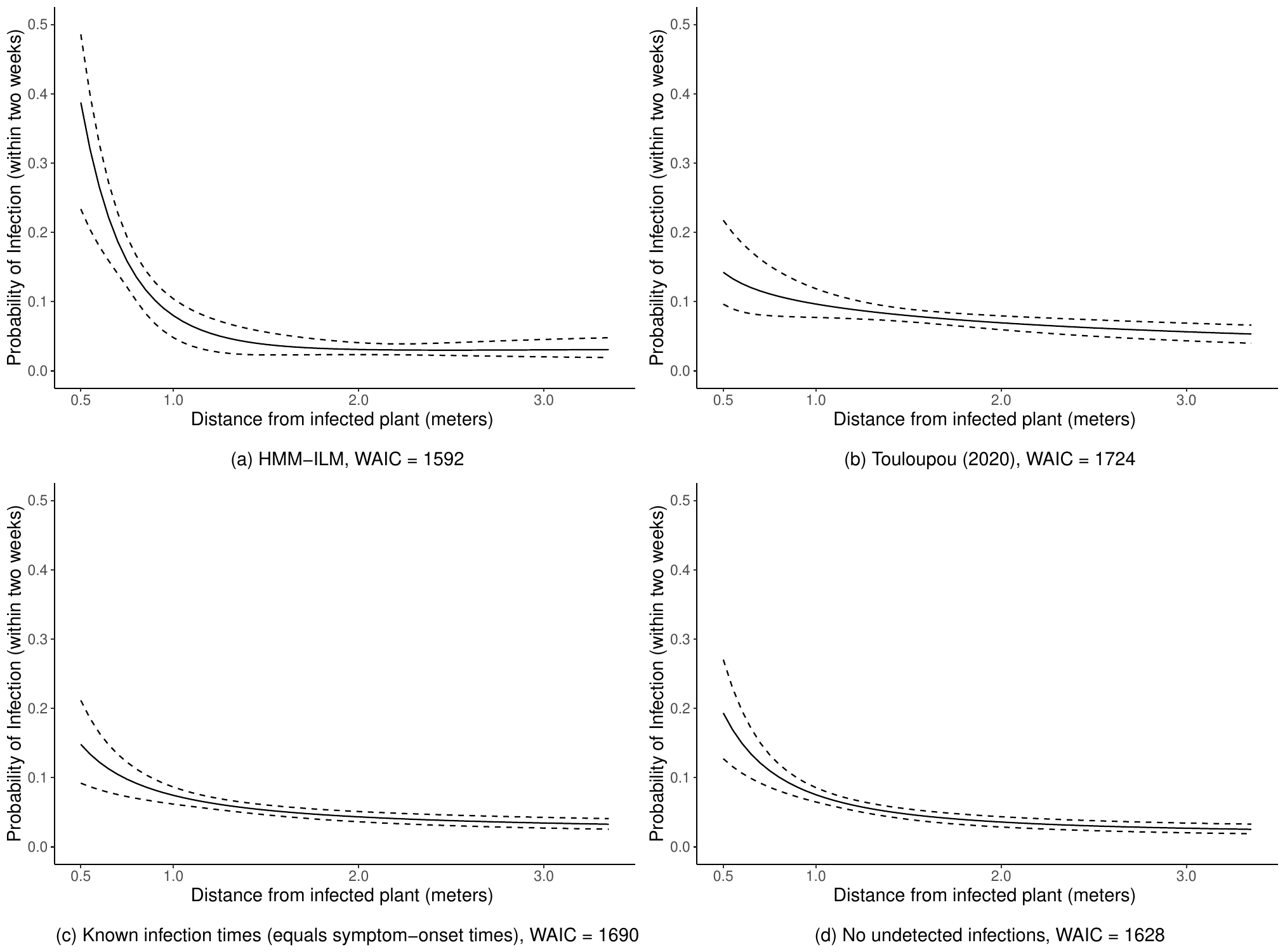}
	\caption{ Shows the probability that a susceptible plant is infected (within two weeks), given a single infectious plant, as a function of the distance between the two plants, for four models fitted to the TSWV experiment. Shows posterior medians and 95\% credible intervals. \label{fig:TSWV_comparison} }
\end{figure}

Figure \ref{fig:TSWV_comparison} compares the probability of infection versus distance and WAIC of (a) {\color{black}the proposed HMM-ILM}, (b) an HMM-ILM with (\ref{eqn:GHMM_obs}) replaced by (\ref{eqn:GHMM_obs_cont}) \citep{touloupou_scalable_2020}, (c) an HMM-ILM with (\ref{eqn:GHMM_obs}) replaced by (\ref{eqn:GHMM_obs_KIT}) (known infection times), and (d) an HMM-ILM fixing $S_{it}=1$ for all {\color{black}plants that did not show symptoms} (no undetected infections). For the comparison models, we used the same priors and also kept the other model specifications the same. As shown in the figure, {\color{black}the} HMM-ILM has the lowest WAIC by a significant amount (more than 5), meaning that it has the best fit to the data. In addition, assuming an {\color{black}independent Bernoulli observation process}, known infection times, or {\color{black}that there were no undetected infections}, all result in estimating a much flatter distance kernel. For example, for a susceptible plant half a meter away from an infectious plant, the HMM-ILM estimates an infection risk of 39\% (23\%, 49\%), while the model assuming that {\color{black}all plants that did not show symptoms escaped infection} estimates an infection risk of only 19\% (12\%, 27\%), which is half as large. A possible explanation for this difference is that, from Figure \ref{fig:TSWV_likely_infections}, many of the plants {\color{black}that did not show symptoms} and were likely infected were {\color{black}adjacent to} plants with known infections in the same row, half a meter away. Therefore, if we assume that none of those {\color{black}symptomless} plants were actually infected, we would expect the effect of disease spread between close plants to be underestimated.

To help dig deeper into the differences between the models, we give the full set of posterior distributions of each model from Figure \ref{fig:TSWV_comparison} in SM Section 5.2. Interestingly, assuming {\color{black}that all plants that did not show symptoms escaped infection} leads to a posterior for $\theta$ that is concentrated close to one. This makes sense intuitively. If we assume that none of the {\color{black}plants that did not show symptoms during the study} were infected, then every plant that was infected must have shown symptoms before the end of the study. This then implies a high chance of {\color{black}a plant showing symptoms.} However, since it takes 2-4 weeks for symptoms to appear in a plant \citep{brust2024}, it seems unlikely that we would have a probability close to one for a plant to {\color{black}show symptoms} in the same biweek it was infected. Also interestingly,  assuming {\color{black}an independent Bernoulli observation model} leads to a posterior for $m$ that is concentrated close to one. Again, this makes sense intuitively. If individuals are infectious for more than one time step, {\color{black}then Equation (\ref{eqn:GHMM_obs_cont}) has a chance of generating multiple symptom-onset times for an individual. Therefore, the only way for this model to explain symptom-onset data} is if $m=1$ ($\theta$ could also be 0, but then no one would show symptoms). However, this seems unlikely given that plants are infectious with TSWV until death \citep{nsw_tomato_spotted_2024}.

\section{Discussion} \label{sec:discussion}

{\color{black}We have proposed an autoregressive coupled hidden Markov model for analyzing individual-level outbreak data. The model can capture many important aspects of an epidemic, including heterogeneous mixing, asymptomatic and late presymptomatic infections (which can be identified), and random incubation and infectious periods. Also, it only requires the symptom-onset times to fit, which are often the only observations available in an outbreak investigation \citep{stockdale_modelling_2017,kypraios_tutorial_2017,o2019markov,stockdale_pair-based_2021,almutiry2021continuous,aristotelous_posterior_2023}. Bayesian inference is straightforward using the iFFBS algorithm, which samples all unknown state indicators for an individual directly from their full conditional distribution. No complex Metropolis-Hastings proposals or reversible jump steps are required for sampling the hidden infectious process, like with past methods \citep{oneill_bayesian_1999,jewell_bayesian_2009}.} 

{\color{black}There are other approaches in the literature that can infer the parameters of an ILM using only symptom-onset times \citep{o2019markov}. However, these approaches have either assumed that the infection or removal times are known and equal to the symptom-onset times \citep{oneill_bayesian_1999,bu_likelihood-based_2022} or they have not accounted for undetected infections (we discuss this more in the next paragraph) \citep{cauchemez_bayesian_2004,almutiry2021continuous}.}
In contrast, our approach does not assume that {\color{black}the onset of symptoms} corresponds to infection or removal. As we showed in our examples, important model results, {\color{black}such as} estimates of $R_0$, can be very sensitive to assumptions about whether the infection or removal times are known. For instance, in our norovirus example, assuming that the removal times were equal to the symptom-onset times led to an unrealistically low estimate of between-nurse transmission. In our TSWV example, assuming that the infection times were equal to the symptom-onset times resulted in estimating a much flatter distance kernel and a worse model fit. 

\cite{almutiry2021continuous} proposed using {\color{black}an} SINR model \citep{jewell_bayesian_2009} to estimate infection and removal times when both are unknown. However, their framework only estimates the infection and removal times of the {\color{black}individuals who showed symptoms} and assumes {\color{black}that any individual who did not show symptoms escaped infection. Therefore, their method cannot account for asymptomatic infections or infections that occur near the end of the study and have not yet developed symptoms, both of which the HMM-ILM can capture.} For the TSWV example, we found that at least 58 of {\color{black}the plants that never showed symptoms} during the experiment were likely infected, {\color{black}an aspect the analysis of} \cite{almutiry2021continuous} {\color{black} did not account for}. {\color{black}Ignoring these undetected infections led to a worse model fit, a much flatter distance kernel, and an unrealistically low estimate of the average length of the incubation period.}

\cite{touloupou_scalable_2020} also analyzed individual-level outbreak data using coupled hidden Markov models. {\color{black}However, their model was designed for laboratory test results collected at predetermined times and should not be used for symptom-onset times. As we showed in Section \ref{sec:TSWV}, their model can produce multiple symptom-onset times for an individual and, therefore, will estimate that individuals are only infectious for one time step.} 

We focused on the case {\color{black}where the time that symptoms first appeared in an individual is observed, since this is almost always reported in an outbreak investigation \citep{becker2017analysis}.} We also showed how the observation model (\ref{eqn:GHMM_obs}) can be modified for when the infection (\ref{eqn:GHMM_obs_KIT}) or removal times (\ref{eqn:GHMM_obs_KRT}) are observed. An important avenue for future work is to consider further observation models. For instance, in the Tristan da Cunha cold example from \cite{becker2017analysis}, they observed the start and end of symptoms for each individual who showed symptoms. This could be modelled by letting $y_{it}|S_{it}=2$ follow a Markov chain. {\color{black}In addition, sometimes the onsets of multiple interacting symptoms are observed in an outbreak investigation, for example, prodrome and rash for measles \citep{neal_statistical_2004}. Therefore, multivariate extensions of Equation (\ref{eqn:GHMM_obs}) could be very useful.} 

There are some important limitations of our work. We only consider SIR compartmental models. The iFFBS algorithm, which is the cornerstone of our inferential procedure, only requires that $S_{it}|\bm{S}_{(-i)(t-1)}$ follow a first-order Markov chain \citep{douwes2024coupled}. Therefore, extensions to more complex compartmental models, such as susceptible-exposed-infectious-removed (SEIR) models \citep{bu_stochastic_2024}, should be straightforward in theory. However, {\color{black}given only symptom-onset times,} SEIR models {\color{black}can} suffer from serious identifiability problems {\color{black}even when making the strong assumption that symptom onset corresponds to removal \citep{o2019markov}. Therefore, additional data and/or informative priors would likely have to be considered to fit such models.} Another limitation is that we assume a geometric distribution for the incubation period. For instance, in the TSWV example, we assume that after a plant becomes infected, it has a constant chance $\theta$ of first showing symptoms each time step. Since it takes 2-4 weeks for the symptoms to develop, it would be more realistic to let $\theta$ be 0 for the first time step of infectiousness, an unknown quantity to be estimated for time step 2, and then 0 again after time step 2. That is, if the plant has not developed symptoms after a month or so, it likely will not develop them. This could be accomplished by splitting up the infectious compartment into sub-compartments. {\color{black}Finally, individuals often change their behaviour during an epidemic in response to others around them showing symptoms \citep{ward2023bayesian,ward2025framework} and to their own symptoms. This could be incorporated into the HMM-ILM by letting $\beta_{j \to i,t}$ depend on $y_{j0},\dots,y_{j(t-1)}$ {\color{black}(i.e., the transmission rate depends on whether the infective has shown symptoms).} Note that the iFFBS algorithm allows for autoregression in the transition probabilities \citep{douwes2024coupled}.} We will consider extensions {\color{black}such as} this in future work.

\section*{Acknowledgments}

Funding for the project was provided by the Canadian Statistical Sciences Institute (CANSSI) Distinguished Postdoctoral Fellowship. A. M. Schmidt and R. Deardon acknowledge financial support from the Natural Sciences and Engineering Research Council (NSERC) of Canada Discovery Grants program (Schmidt—RGPIN/2024-04312, Deardon—RGPIN/03292-2022).

\bibliography{HMMILM}

\end{document}


\maketitle

\tableofcontents

\newpage

\section{The Individual Forward Filtering Backward Sampling (iFFBS) Algorithm}  

In this section, we will go into more detail concerning the hybrid Gibbs sampling algorithm we use to draw from the joint posterior of $\bm{S}$ and $\bm{v}=(\theta,m,\alpha,\bm{\beta})^T$, $p(\bm{S},\bm{v}|\bm{y})$. We will borrow all notation from the main text. We first choose initial values for the Gibbs sampler, $\bm{S}^{[0]}$ and $\bm{v}^{[0]}$. The initial values for the parameter vector $\bm{v}$, $\bm{v}^{[0]}$, are generated randomly within the parameter space. To ensure valid initial values for the state indicators, we first set  $S_{it}^{[0]}=1$ for all {\color{black}individuals $i$ who did not show symptoms} and for all $t$. Then, for an individual $i$ {\color{black}with symptom onset at time $k$}, we let $S_{it}^{[0]}=1$ for $t<k$,  $S_{it}^{[0]}=2$ for $t=k$, and $S_{it}^{[0]}=3$ for $t>k$ (if $y_{ik}=1$ then $S_{ik}^{[0]}$ must be 2). If the observation process, Equation {\color{black}(4)} in the main text, is changed, then we modify $\bm{S}^{[0]}$ accordingly to ensure $p\left(\bm{y},\bm{S}^{[0]}|\bm{v}^{[0]}\right)>0$; see Equation (5) of the main text. 

After setting the initial values, the following steps are repeated for $q=1,\dots,Q$, where $Q$ is the total number of iterations for the Gibbs sampler,
\begin{enumerate}
    \item Sample $\bm{v}^{[q]}$ from $p(\bm{v}|\bm{S}^{[q-1]},\bm{y})$.
    \item Sample $\bm{S}_{i(0:T)}^{[q]}$ from $p(\bm{S}_{i(0:T)}|\bm{S}_{1(0:T)}^{[q]},\dots,\bm{S}_{i-1(0:T)}^{[q]},\bm{S}_{i+1(0:T)}^{[q-1]},\dots,\bm{S}_{N(0:T)}^{[q-1]},\bm{y},\bm{v}^{[q]})$ for $i=1,\dots,N$.
    \end{enumerate}

As mentioned in the main text, step 1 is mainly broken up into slice or Metropolis-Hastings steps. Here, we will provide the iFFBS algorithm needed to sample from $p(\bm{S}_{i(0:T)}|\bm{S}_{(-i)(0:T)},\bm{y},\bm{v})$ for step 2. The algorithm was originally proposed by \cite{touloupou_scalable_2020}. We will sometimes use the subscript $t_1:t_2$ to denote a temporally indexed vector subset to the interval $t_1$ to $t_2$, e.g., $y_{i(0:t)}=(y_{i0},\dots,y_{it})^{T}$. 

First note that,
\begin{align} \label{eqn:iFFBS1}
p(\bm{S}_{i(0:T)}|\bm{S}_{(-i)(0:T)},\bm{y},\bm{v}) &= p(S_{iT}|\bm{S}_{(-i)(0:T)},\bm{y},\bm{v}) \prod_{t=0}^{T-1} p(S_{it}|S_{i(t+1)},\bm{S}_{(-i)(0:t+1)},\bm{y}_{i(0:t)},\bm{v}).
\end{align} Therefore, to sample $\bm{S}_{i(0:T)}^{[q]}$ from $p(\bm{S}_{i(0:T)}|\bm{S}_{(-i)(0:T)},\bm{y},\bm{v})$, we can sample $S_{iT}^{[q]}$ from \newline $p(S_{iT}|\bm{S}_{(-i)(0:T)},\bm{y},\bm{v})$, then $S_{i(T-1)}^{[q]}$ from $p(S_{i(T-1)}|S_{iT}^{[q]},\bm{S}_{(-i)(0:T)},\bm{y}_{i(0:T-1)},\bm{v})$, and so on. Now note that, from Bayes' Theorem, 
\begin{align}\label{eqn:iFFBS2}
p(S_{it}|S_{i(t+1)},\bm{S}_{(-i)(0:t+1)},\bm{y}_{i(0:t)},\bm{v}) \propto p(S_{i(t+1)}|S_{it},\bm{S}_{(-i)(0:t)},m,\alpha,\bm{\beta})p(S_{it}|\bm{S}_{(-i)(0:t+1)},\bm{y}_{i(0:t)},\bm{v}).
\end{align} The first term on the right-hand side of (\ref{eqn:iFFBS2}), $p(S_{i(t+1)}|S_{it},\bm{S}_{(-i)(0:t)},m,\alpha,\bm{\beta})$, can be derived from the transition matrix in Equation {\color{black}(1)} of the main text. The probabilities \newline $P(S_{it}=s|\bm{S}_{(-i)(0:t+1)},\bm{y}_{i(0:t)},\bm{v})$ for $s=1,2,3$ and $t=0,\dots,T-1$ and $P(S_{iT}=s|\bm{S}_{(-i)(0:T)},\bm{y},\bm{v})$ for $s=1,2,3$ are known as the filtered probabilities. These are calculated iteratively using the forward part of the algorithm from $t=0$.

Starting at $t=0$ we have,
\begin{align} \label{eqn:iFFBS3}
\begin{split}
p(S_{i0}|\bm{S}_{(-i)(0:1)},y_{i0},\bm{v}) &\propto p(S_{i0})p(\bm{S}_{(-i)1}|\bm{S}_{(-i)0},S_{i0},m,\alpha,\bm{\beta}) \\
&\propto p(S_{i0}) \prod_{j:i \in NE(j)} p(S_{j1}|S_{j0},\bm{S}_{
(-j)0},m,\alpha,\bm{\beta}).
\end{split}
\end{align} Here, $p(S_{i0})$ is the initial state distribution, which is set by the modeler, and \newline $p(S_{j1}|S_{j0},\bm{S}_{
(-j)0},m,\alpha,\bm{\beta})$ represents a transition probability of the Markov chain for individual $j$.
Since $p(S_{j1}|S_{j0},\bm{S}_{
(-j)0},m,\alpha,\bm{\beta})$ only depends on whether individual $i$ is infectious, only two values of $\prod_{j:i \in NE(j)} p(S_{j1}|S_{j0},\bm{S}_{
(-j)0},m,\alpha,\bm{\beta})$ need to be calculated. Also, since $S_{i0}$ can only take three values, it is straightforward to derive the filtered probabilities from (\ref{eqn:iFFBS3}),
\begin{align} \label{eqn:iFFBS4}
\begin{split}
P(S_{i0}=s|\bm{S}_{(-i)(0:1)},y_{i0},\bm{v})=\frac{P(S_{i0}=s)\prod\limits_{\substack{j \, : \, i \,\in \, NE(j) \\ S_{i0}=s}}p(S_{j1}|S_{j0},\bm{S}_{
(-j)0},m,\alpha,\bm{\beta})}{\sum_{k=1}^{3}P(S_{i0}=k)\prod\limits_{\substack{j \, : \, i \,\in \, NE(j) \\ S_{i0}=k}}p(S_{j1}|S_{j0},\bm{S}_{
(-j)0},m,\alpha,\bm{\beta})},
\end{split}
\end{align} for $s=1,2,3$.

For $t=1,\dots,T-1$, we have 
\begin{align*}
p(S_{it}|\bm{S}_{(-i)(0:t+1)},\bm{y}_{i(0:t)},\bm{v}) &\propto p(y_{it}|S_{it},\bm{y}_{i(0:t-1)},\theta)p(S_{it}|\bm{S}_{(-i)(0:t)},\bm{y}_{i(0:t-1)},\bm{v}) \\[5pt]
& \,\,\,\, \times\prod_{j:i \in NE(j)} p(S_{j(t+1)}|S_{jt},\bm{S}_{
(-j)t},m,\alpha,\bm{\beta}).
\end{align*} Here{\color{black},} $p(y_{it}|S_{it},y_{i(0:t-1)},\theta)$ is given by the observation model from Equation {\color{black}(4)} in the main text. The probabilities $P(S_{it}=s|\bm{S}_{(-i)(0:t)},\bm{y}_{i(0:t-1)},\bm{v})$ for $s=1,2,3$ are known as the predictive probabilities and can be calculated as follows,
\begin{align*}
P(S_{it}&=s|\bm{S}_{(-i)(0:t)},\bm{y}_{i(0:t-1)},\bm{v}) = \\ &\,\,\sum_{k=1}^{3}P(S_{it}=s|S_{i(t-1)}=k,\bm{S}_{(-i)(t-1)},m,\alpha,\bm{\beta})P(S_{i(t-1)}=k|\bm{S}_{(-i)(0:t)},\bm{y}_{i(0:t-1)},\bm{v}),
\end{align*} where $P(S_{i(t-1)}=k|\bm{S}_{(-i)(0:t)},\bm{y}_{i(0:t-1)},\bm{v})$ for $k=1,2,3$ are the previous filtered probabilities. It then follows that,
\begin{align} \label{eqn:iFFBS5}
\begin{split}
&P(S_{it}=s|\bm{S}_{(-i)(0:t+1)},\bm{y}_{i(0:t)},\bm{v})= \\
&\frac{p(y_{it}|S_{it}=s,\bm{y}_{i(0:t-1)},\theta)P(S_{it}=s|\bm{S}_{(-i)(0:t)},\bm{y}_{i(0:t-1)},\bm{v})\prod\limits_{\substack{j \, : \, i \,\in \, NE(j) \\ S_{it}=s}}p(S_{j(t+1)}|S_{jt},\bm{S}_{
(-j)t},m,\alpha,\bm{\beta})}{\sum_{k=1}^{3}p(y_{it}|S_{it}=k,\bm{y}_{i(0:t-1)},\theta)P(S_{it}=k|\bm{S}_{(-i)(0:t)},\bm{y}_{i(0:t-1)},\bm{v})\prod\limits_{\substack{j \, : \, i \,\in \, NE(j) \\ S_{it}=k}}p(S_{j(t+1)}|S_{jt},\bm{S}_{
(-j)t},m,\alpha,\bm{\beta})},
\end{split}
\end{align} for $s=1,2,3$.

Calculating the filtered probabilities for $t=T$ is similar except there are no forward product terms,
\begin{align}  \label{eqn:iFFBST}
\begin{split}
P(S_{iT}=s|\bm{S}_{(-i)(0:T)},\bm{y},\bm{v})= 
\frac{p(y_{iT}|S_{iT}=s,\bm{y}_{i(0:T-1)},\theta)P(S_{iT}=s|\bm{S}_{(-i)(0:T)},\bm{y}_{i(0:T-1)},\bm{v})}{\sum_{k=1}^{3}p(y_{iT}|S_{iT}=k,\bm{y}_{i(0:T-1)},\theta)P(S_{iT}=k|\bm{S}_{(-i)(0:T)},\bm{y}_{i(0:T-1)},\bm{v})},
\end{split}
\end{align} for $s=1,2,3$.

Once the filtered probabilities have been calculated $\bm{S}_{i(0:T)}$ can be sampled backwards using Equations (\ref{eqn:iFFBS1}) and (\ref{eqn:iFFBS2}). Firstly, $S_{iT}^{[q]}$ is sampled from $p(S_{iT}|\bm{S}_{(-i)(0:T)},\bm{y},\bm{v})$ using (\ref{eqn:iFFBST}). Then, for $t=T-1,\dots,0$, $S_{it}^{[q]}$ is drawn from the density defined by,
\begin{align*}
&P(S_{it}=s|S_{i(t+1)}=S_{i(t+1)}^{[q]},\bm{S}_{(-i)(0:t+1)},\bm{y}_{i(0:t)},\bm{v}) = \\
&\frac{P(S_{i(t+1)}=S_{i(t+1)}^{[q]}|S_{it}=s,\bm{S}_{(-i)t},m,\alpha,\bm{\beta})P(S_{it}=s|\bm{S}_{(-i)(0:t+1)},\bm{y}_{i(0:t)},\bm{v})}{\sum_{k=1}^{3} P(S_{i(t+1)}=S_{i(t+1)}^{[q]}|S_{it}=k,\bm{S}_{(-i)t},m,\alpha,\bm{\beta})P(S_{it}=k|\bm{S}_{(-i)(0:t+1)},\bm{y}_{i(0:t)},\bm{v})},
\end{align*} for $s=1,2,3$. Here, $P(S_{it}=s|\bm{S}_{(-i)(0:t+1)},\bm{y}_{i(0:t)},\bm{v})$ is given by either Equation (\ref{eqn:iFFBS5}) or (\ref{eqn:iFFBS4}) (for t=0).

In NIMBLE, we code the iFFBS algorithm on the log scale to improve numerical stability. All NIMBLE R code for the custom iFFBS samplers is provided on GitHub {\color{black}(\url{https://github.com/Dirk-Douwes-Schultz/HMM_ILM_Code})}. Note that the only calculations that separate the iFFBS algorithm from a traditional FFBS algorithm for hidden Markov models \citep{chibCalculatingPosteriorDistributions1996,fruhwirth-schnatterFiniteMixtureMarkov2006} are the forward product terms $\prod_{j:i \in NE(j)} p(S_{j(t+1)}|S_{jt},\bm{S}_{
(-j)t},m,\alpha,\bm{\beta})$, which are needed to account for between chain dependencies.

\section{Simulation Study to Assess Parameter Recovery} \label{sec:SM_sim}

We designed a simulation study to ensure that our hybrid Gibbs sampling algorithm could recover the true parameter values of the HMM-ILM. We simulated repeatedly from an HMM-ILM specified as in the tomato spotted wilt virus (TSWV) example in Section 4.2 of the main text. {\color{black}As} in that example, we assumed $N=520$ individuals in a 26-by-20 grid, with a meter between rows and half a meter between individuals within the same row (see Figure {\color{black}5} in the main text). We assumed that the individuals were observed over $t=1,\dots, T=7$ time periods and that at $t=0$, each individual had a 1\% chance of being initially infected. We specified the effect of disease spread from individual $j$ to individual $i$, see Equations (2)-(3) in the main text, as
\begin{align*} 
\begin{split}
\beta_{j\to i,t}= \beta_{j\to i}=\beta_0d_{ij}^{-a}\left(1-\ln(d_{ij})(\beta_1-a)+0.5\ln(d_{ij})^2(\beta_1-a)^2\right),
\end{split}
\end{align*} where $d_{ij}$ is the distance between the two individuals and $a=1.35$. Finally, we let $NE(i)$, from Equation {\color{black}(2)} of the main text, include all individuals at most three individuals away from individual $i$, including on the diagonal. See Section 4.2 of the main text for more details on how the above specifications were determined.

For the true parameter values, we took the posterior medians estimated in Section 4.2 of the main text: $\theta=0.55$, $m=16$, $\alpha=0.015$, $\beta_0=0.07$, and $\beta_1=3$. The values for $\theta$ and $m$ mean that if an individual becomes infected, they have a 55\% chance of {\color{black}first showing symptoms} each time step and that the infectious period lasts an average of 16 time steps. The values for $\alpha$, $\beta_0$, and $\beta_1$ yield a fairly short range of dispersal for the disease; see Figure {\color{black}6} in the main text. We simulated from the above HMM-ILM 232 times {\color{black}(originally we ran 250 simulations; however, 18 did not finish within their allotted time {\color{black}because the cluster nodes became unresponsive, which is unrelated to our model or code)}}. Figure \ref{fig:SM_sim_ex} below shows one of the simulated outbreaks. 

\begin{figure}[p!]
 	\centering
 	\includegraphics[width=\textwidth]{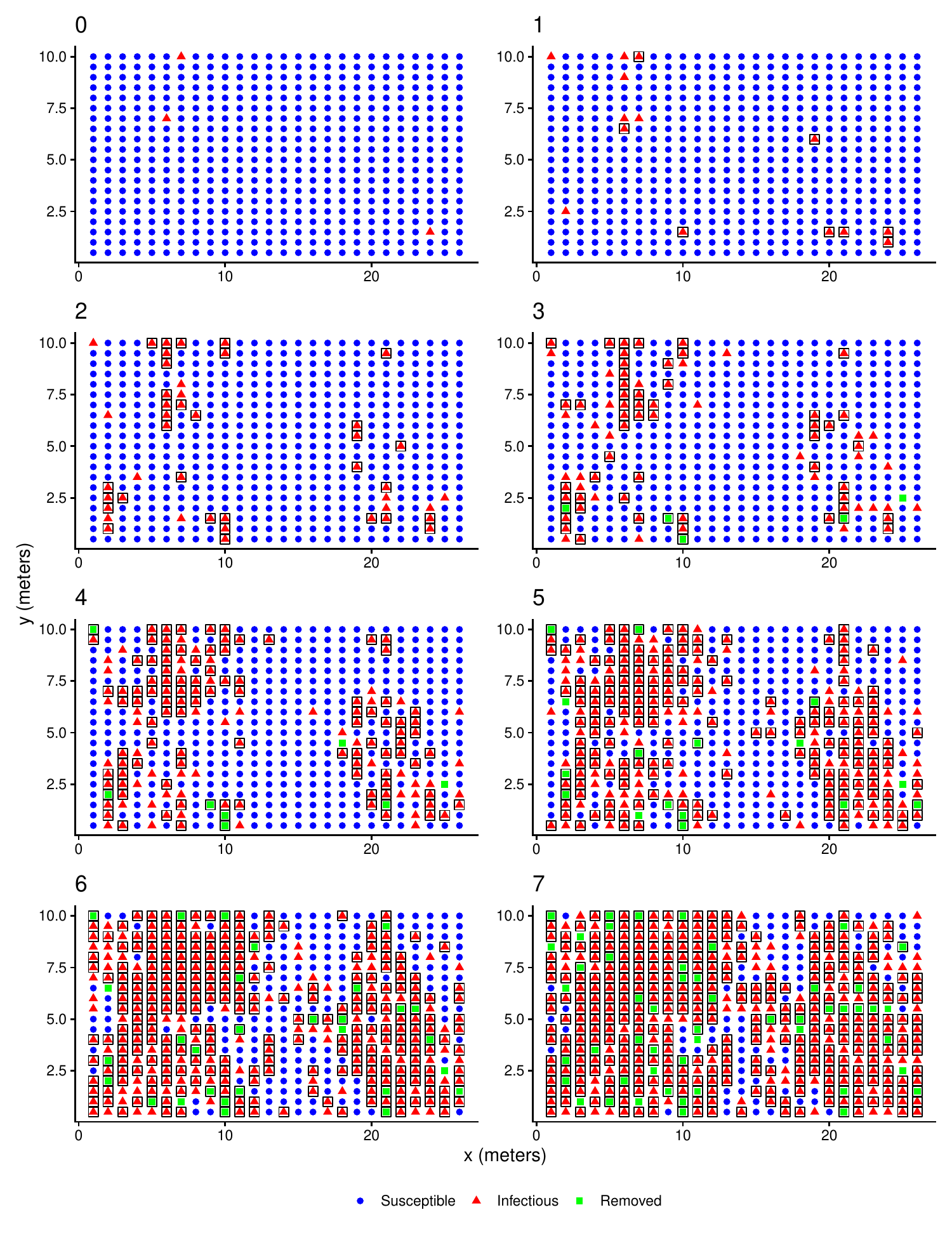}
	\caption{ Shows one of the simulated outbreaks from the simulation study in SM Section 2 with $m=16$. A box is drawn around individuals who have {\color{black}shown symptoms}. From the final plot, there are 83 undetected infections in total (i.e., 83 red triangles or green squares without boxes around them).\label{fig:SM_sim_ex} }
\end{figure}

We fit a correctly specified HMM-ILM to each of the 232 simulated outbreaks using our hybrid Gibbs sampling algorithm. We used the same {\color{black} independent} prior distributions as in Section 4.2 of the main text: $m\sim \text{Unif(1,20)}$, $\theta \sim \text{Unif}(0,1)$, $\alpha \sim \text{Unif}(0,1)$, $\beta_0 \sim \text{Unif}(0,1)$, and $\beta_1 \sim \text{Unif}(0,20)$. {\color{black}As} in the main text, we ran our hybrid Gibbs sampling algorithm for 200,000 iterations on three chains with an initial burn-in of 50,000 iterations. We placed an automated factor slice sampler \citep{tibbits2014automated} on the entire parameter vector $\bm{v}=(m,\theta,\alpha,\beta_0,\beta_1)^T$. Since we cannot visually assess the convergence for all 232 fits, we wanted to ensure the best possible mixing for $\bm{v}$. For each fit, convergence was checked using the minimum effective sample size ($>1000$) and the maximum Gelman-Rubin statistic ($<1.05$) \citep{plummerCODAConvergenceDiagnosis2006}. Our Gibbs sampler passed the convergence checks for $230/232=99.14\%$ of the simulations, illustrating a good convergence rate.

\begin{figure}[t]
 	\centering
 	\includegraphics[width=.8\textwidth]{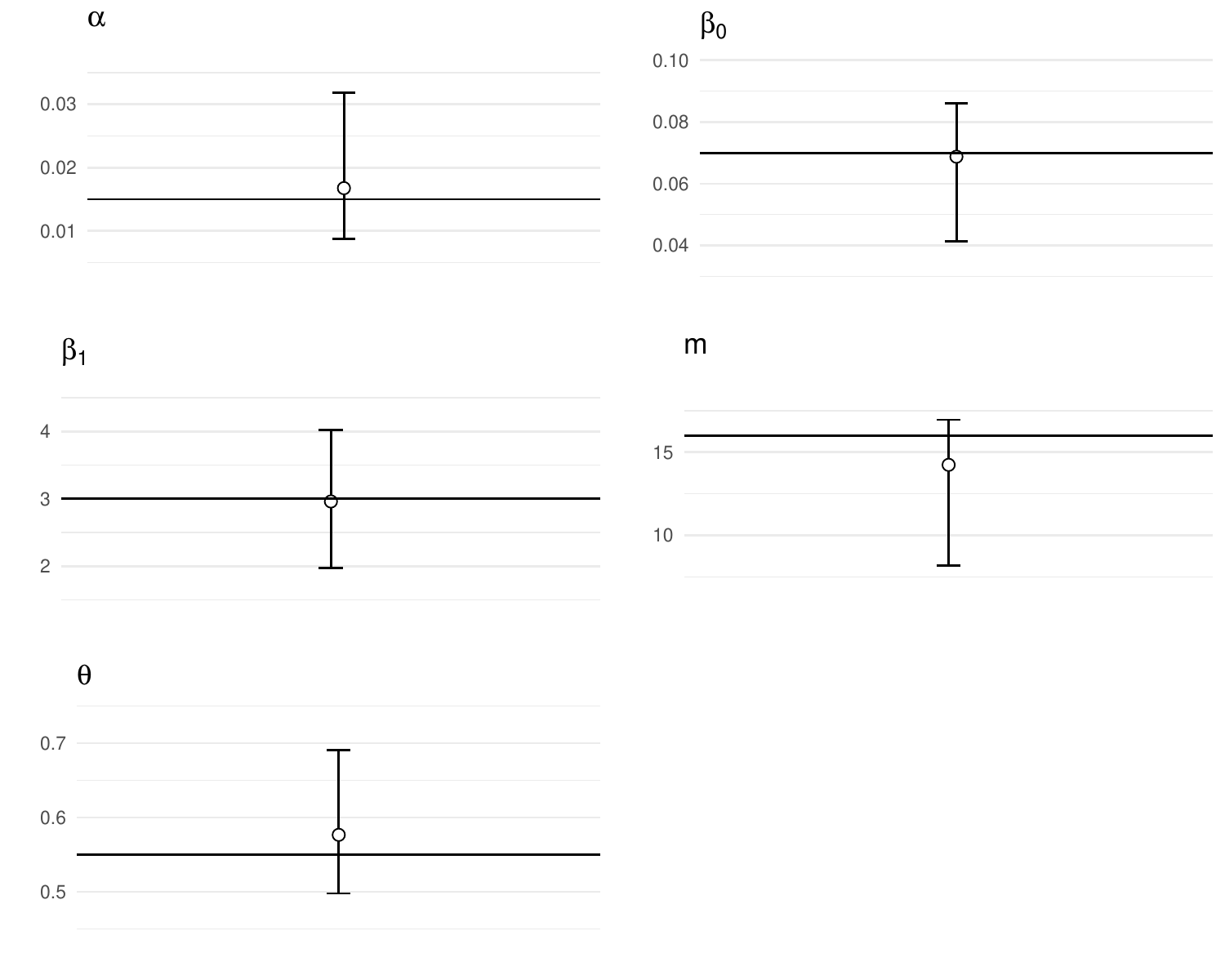}
	\caption{ Shows the sample medians (circles) and 95\% quantile intervals (.025 and .975 quantiles) (caps) of the posterior medians from fitting 230 replications of the HMM-ILM described in SM Section 2 with m=16. The horizontal lines are drawn at the true parameter values.\label{fig:SM_sim_medians} }
\end{figure}

\begin{figure}[t]
 	\centering
 	\includegraphics[width=\textwidth]{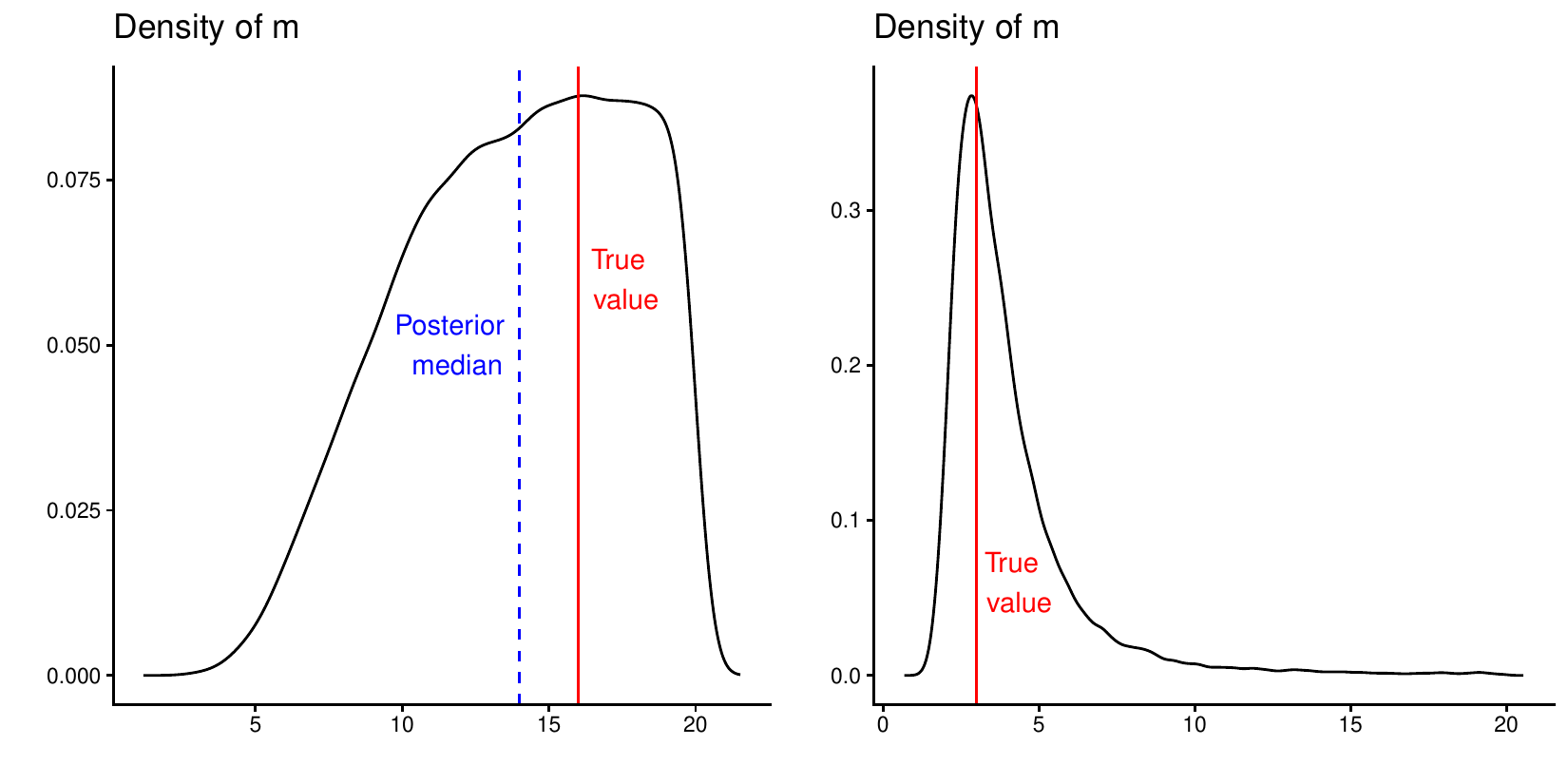}
	\caption{ Shows the posterior distribution of $m$ from fitting a single simulation of the HMM-ILM described in SM Section 2 with $m=16$ (left) and $m=3$ (right).\label{fig:SM_sim_m_compare} }
\end{figure}

Figure \ref{fig:SM_sim_medians} shows the {\color{black}sample} medians and 95\% quantile intervals (0.025 and 0.975 quantiles) of the 230 posterior medians from the converged fits. The horizontal lines are drawn at the true parameter values. From the figure, the posteriors of $\alpha$, $\beta_0$, and $\beta_1$ are all centered very close to the true parameter values on average. The posteriors of $\theta$ and $m$ are centered slightly off the true parameter values on average (the sample medians of the 230 posterior medians are .58 and 14.24 versus true values of .55 and 16, respectively). However, the discrepancy is not large enough to dramatically change any conclusions {\color{black} and seems to result from relatively flat marginal posterior distributions;} the left-hand graph of Figure \ref{fig:SM_sim_m_compare} shows the posterior distribution of $m$ from fitting one of the simulations. 

Due to the flatness in the posterior, the left-hand plot of Figure \ref{fig:SM_sim_m_compare} suggests it may be difficult to estimate $m$ when the true value is greater than $T$, the length of the observation period. We also observed this flatness in the posterior of $m$ for the TSWV example in Section 4.2 of the main text. For sensitivity analysis, we ran/fit an additional simulation with $m=3$ instead of $16$. The posterior distribution of $m$ for this additional simulation is given in the right-hand plot of Figure \ref{fig:SM_sim_m_compare}. From the plot, the posterior of $m$ is much more concentrated around the true parameter value when the true value is less than $T$ compared to when the true value is greater than $T$.

Table \ref{tab:SM_tab1} shows the coverage and average width of the 95\% credible intervals. The average coverage of the 95\% credible intervals is 95.91\%, which is close to the nominal values. However, the coverage of $m$ is relatively high at $99.56\%$. This is likely due to the truncation of the prior distribution, $m \sim \text{Unif}(1,20)$. For sensitivity analysis, we refit each of the 230 replications using an unbounded prior for $m$, $1/m \sim \text{Unif}(0,1)$ (or $p(m)=m^{-2}I[m>1]$). Our Gibbs sampler converged for all but 2 of the 230 simulations when using the new prior. The results for $\alpha$, $\beta_0$, $\beta_1$, and $\theta$ were all very similar to those shown in Table \ref{tab:SM_tab1} and Figure \ref{fig:SM_sim_medians}, so we will not report them here. The coverage of the 95\% credible interval for $m$ went from 99.56\% to 96.26\%. However, the average width of the credible interval went from 13.03 to 211.55. Indeed, the 95\% credible interval for $m$ often included values as high as 80 years or so, which is not realistic as the length of the infectious period. Therefore, some upper bound on $m$ seems necessary, and we recommend using a $\text{Unif}(0,b)$ prior {\color{black}as} we do in the main text. Note, having the prior for $m$ quickly approach 0 does not appear to be enough to prevent the large width of the credible interval, since $p(m)=m^{-2}I[m>1]$ does rapidly converge to 0.

\renewcommand{\arraystretch}{1.5}
\begin{table}[t]
\centering
\caption{Shows the coverage and average width of the 95\% credible intervals (CIs) from the simulation study in SM Section 2 with $m=16$ and prior $m\sim \text{Unif}(1,20)$.}
\label{tab:SM_tab1}
\begin{tabular}{@{}lcc@{}}
\toprule
\textbf{Parameter} & \textbf{Coverage of 95\% CI (\%)} & \textbf{Average Width of 95\% CI} \\ \midrule
$\alpha$           & 93.91                  & 0.02                    \\
$\beta_0$          & 95.22                  & 0.04                    \\
$\beta_1$          & 95.65                  & 2.00                     \\
$m$                & 99.56                  & 13.03                  \\
$\theta$           & 95.21                  & 0.23                    \\ \bottomrule
\end{tabular}
\end{table}

\begin{figure}[]
 	\centering
 	\includegraphics[width=\textwidth]{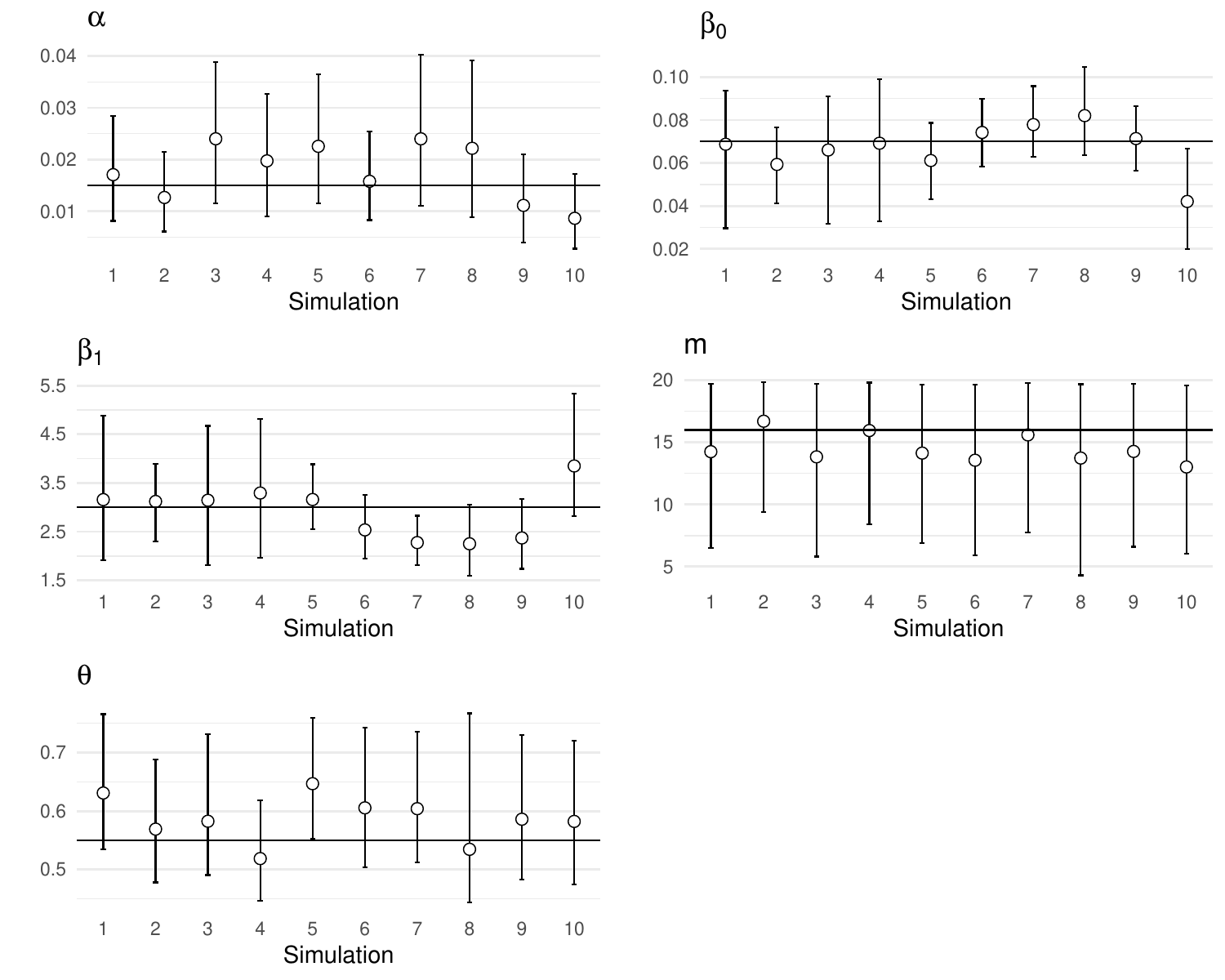}
	\caption{ Shows the posterior medians (circles) and 95\% credible intervals (caps) from fitting the first 10 replications of the HMM-ILM described in SM Section 2 with m=16 and prior $m\sim \text{Unif}(1,20)$. The horizontal lines are drawn at the true parameter values.\label{fig:SM_sim_medians_CIs} }
\end{figure}

{\color{black}Finally, to help visualize the individual simulation results, Figure \ref{fig:SM_sim_medians_CIs} shows the posterior medians and 95\% credible intervals from fitting the first 10 replications. Interestingly, there is considerable variation in the widths of the credible intervals across simulations. This shows that even with the same underlying true parameter values, different outbreaks can carry vastly different amounts of information about the parameters.}

\section{Widely Applicable Information Criterion (WAIC)}

The widely applicable information criterion (WAIC) is a popular model selection criterion for Bayesian models \citep{gelman2014understanding,reich2019bayesian}. However, it is not immediately clear how to calculate the WAIC for a state-space model due to the complex dependency structure of such models \citep{auger-methe_guide_2021}. One possibility is to calculate the WAIC conditional on the underlying latent states \citep{auger-methe_guide_2021,kreuzer2022bayesian}, which for the HMM-ILM is equivalent to,
\begin{align} \label{eqn:WAIC_cond}
\begin{split}
\text{lpdd}^{\text{cond}} & = \sum_{i=1}^{N} \sum_{t=1}^{T}\log\left(\frac{1}{Q-M}\sum_{q=M+1}^{Q}p\left(y_{it}|S_{it}^{[q]},y_{i0},\dots,y_{i(t-1)},\theta^{[q]}\right)\right), \\
\text{pwaic}^{\text{cond}} &= \sum_{i=1}^{N} \sum_{t=1}^{T} \text{Var}_{q=M+1}^{Q} \log\left(p\left(y_{it}|S_{it}^{[q]},y_{i0},\dots,y_{i(t-1)},\theta^{[q]}\right)\right), \\
\text{WAIC}^{\text{cond}} &= -2\left(\text{lpdd}^{\text{cond}}-\text{pwaic}^{\text{cond}}\right),
\end{split}
\end{align} where the superscript $[q]$ denotes a draw from the posterior distribution of the variable, $Q$ is the total number of MCMC samples, $M$ is the size of the burn-in sample, and $\text{Var}$ denotes the sample variance. However, as {\color{black}discussed} by \cite{auger-methe_guide_2021}, the conditional WAIC often fails to select the true model, and they recommend marginalizing out the latent states when calculating the WAIC.
For the HMM-ILM, this is equivalent to,
\begin{align} \label{eqn:WAIC_marg}
\begin{split}
\text{lpdd}^{\text{marg}} & = \sum_{t=1}^{T}\log\left(\frac{1}{Q-M}\sum_{q=M+1}^{Q}p\left(\bm{y}_{t}|\bm{y}_{0},\dots,\bm{y}_{t-1},\bm{v}^{[q]}\right)\right), \\
\text{pwaic}^{\text{marg}} &= \sum_{t=1}^{T} \text{Var}_{q=M+1}^{Q} \log\left(p\left(\bm{y}_{t}|\bm{y}_{0},\dots,\bm{y}_{t-1},\bm{v}^{[q]}\right)\right), \\
\text{WAIC}^{\text{marg}} &= -2\left(\text{lpdd}^{\text{marg}}-\text{pwaic}^{\text{marg}}\right).
\end{split}
\end{align} However, calculating $p\left(\bm{y}_{t}|\bm{y}_{0},\dots,\bm{y}_{t-1},\bm{v}\right)$ requires running Hamilton's forward filter \citep{fruhwirth-schnatterFiniteMixtureMarkov2006}, which requires matrix multiplication with a $3^{N}$ by $3^{N}$ matrix \citep{douwes-schultzZerostateCoupledMarkov2022a}. Therefore, it is not computationally feasible to calculate (\ref{eqn:WAIC_marg}).

To deal with the above issues, we follow \cite{douwes2025three} and use a partially marginalized version of the WAIC. The idea is to marginalize as much of $\bm{S}$ as possible out of $p\left(y_{it}|S_{it},y_{i0},\dots,y_{i(t-1)},\theta\right)=p\left(y_{it}|\bm{S},y_{i0},\dots,y_{i(t-1)},\theta\right)$ in Equation (\ref{eqn:WAIC_cond}). Note that from Equation (\ref{eqn:iFFBS5}) above, we can run the forward part of the iFFBS algorithm to calculate,
\begin{align}\label{eqn:WAIC_pm}
p(y_{it}|\bm{S}_{(-i)(0:t)},y_{i0},\dots,y_{i(t-1)},\bm{v}) &= \sum_{k=1}^{3}p(y_{it}|S_{it}=k,\bm{y}_{i(0:t-1)},\theta)P(S_{it}=k|\bm{S}_{(-i)(0:t)},\bm{y}_{i(0:t-1)},\bm{v}).
\end{align} That is, we can marginalize all the states for individual $i$ and all future states for the other individuals. Therefore, to calculate the WAIC, we use $p\left(y_{it}|\bm{S}_{(-i)(0:t)}^{[q]},y_{i0},\dots,y_{i(t-1)},\bm{v}^{[q]}\right)$ in place of $p\left(y_{it}|S_{it}^{[q]},y_{i0},\dots,y_{i(t-1)},\theta^{[q]}\right)$ in Equation (\ref{eqn:WAIC_cond}). It was shown in \cite{douwes2025three}, through a simulation study, that the WAIC calculated in this manner can distinguish between different coupled hidden Markov models.

Note that $p(y_{it}|\bm{S}_{(-i)(0:t)},y_{i0},\dots,y_{i(t-1)},\bm{v})$ in Equation (\ref{eqn:WAIC_pm}) has a convenient interpretation. It represents the distribution of $y_{it}$ given knowledge of the disease states of all other individuals through time $t$. Therefore, models that are effective at borrowing information between individuals will have lower WAICs, which {\color{black}would seem to be} a desirable property for an epidemic model.

\newpage

\section{Further Analysis of the Norovirus Example}

\subsection{State estimates}

\begin{figure}[p]
 	\centering
 	\includegraphics[width=\textwidth]{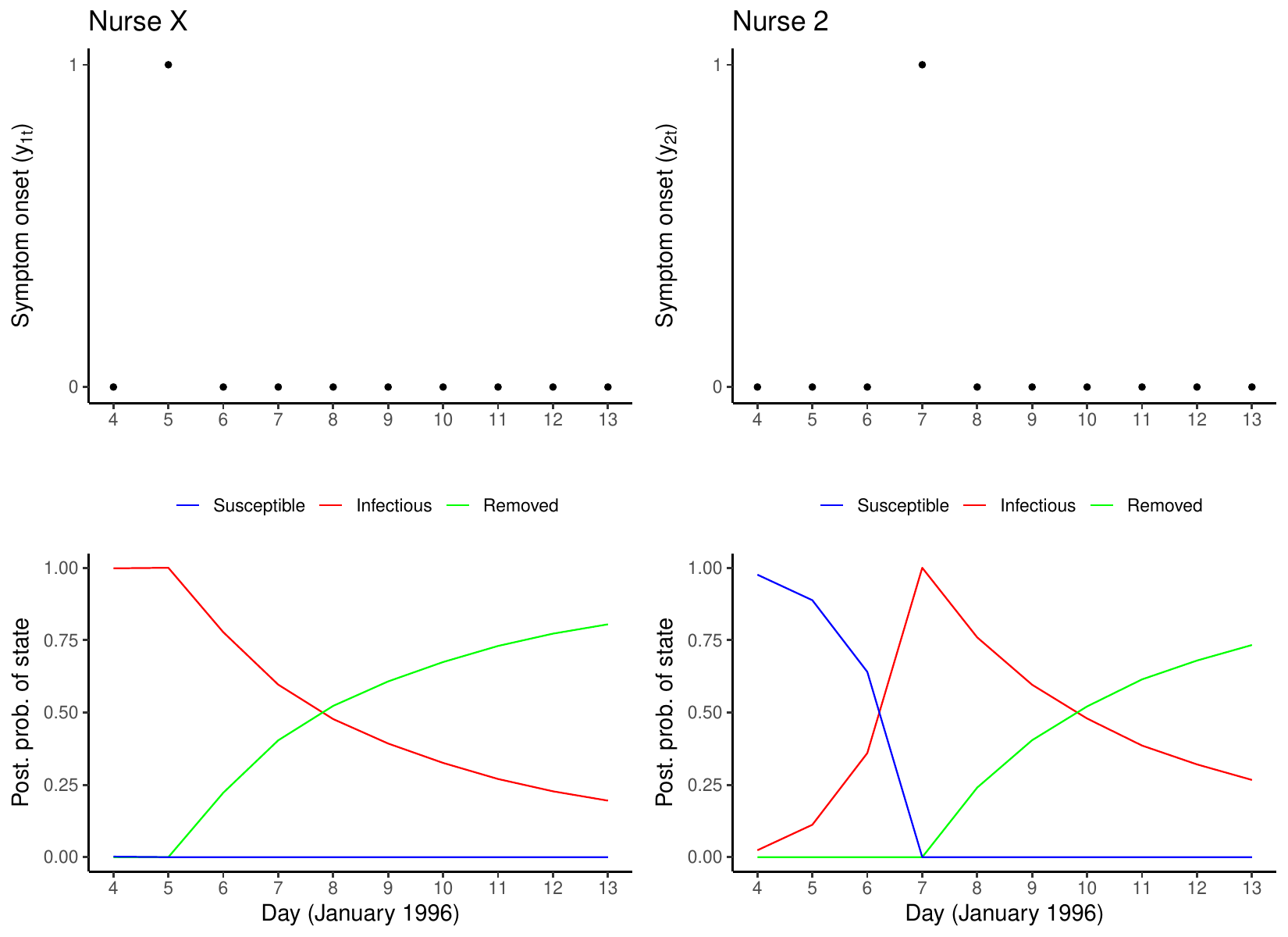}
	\caption{ The top graphs show the day the nurse first reported showing symptoms. The bottom graphs show the posterior probability that the nurse was in the susceptible (blue), infectious (red), and removed (green) states during each day. Nurse X refers to the first case on the fifth. Nurse 2 refers to one of the four nurses {\color{black}who first showed symptoms} on the seventh.\label{fig:SM_detected_noro_se} }
\end{figure}

The posterior probability that nurse $i$ was susceptible ($s=1$), infectious ($s=2$), or removed ($s=3$) on day $t$ can be approximated by $P(S_{it}=s|\bm{y}) \approx \frac{1}{Q-M}\sum_{q=M+1}^{Q}I[S_{it}^{[q]}=s]$ for $i=1,\dots, 89$ and $t=0,\dots,9$. By viewing plots of $P(S_{it}=s|\bm{y})$ versus $t$ for various nurses, we can better understand how the outbreak evolved over time. For example, Figure \ref{fig:SM_detected_noro_se} shows the state estimates for Nurse X (Nurse 1) and Nurse 2 (one of the four nurses {\color{black}who first showed symptoms} on the seventh). From the figure, {\color{black}we can see that} Nurse X was likely infectious at the beginning of the outbreak (i.e., was the index case) and not removed until at least the seventh or eighth. Nurse 2 was likely infected on the fifth or sixth and not removed until at least the ninth or tenth. Examining plots of the other nurses {\color{black}who showed symptoms} revealed a similar pattern of at least 2-4 days from detection to removal and 1-2 days between infection and {\color{black}symptom onset}. 

\begin{figure}[p]
 	\centering
 	\includegraphics[width=.7\textwidth]{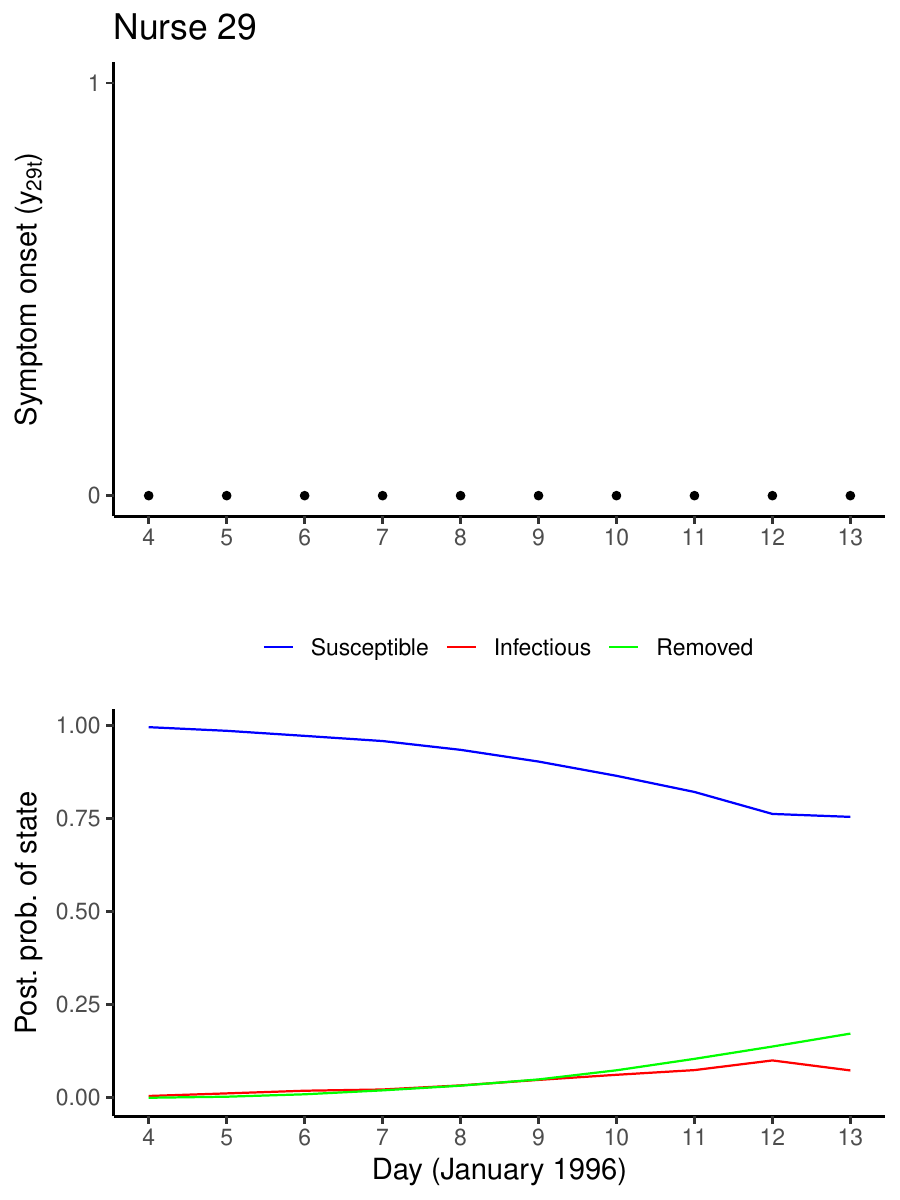}
	\caption{ The top graph shows the day the nurse first reported showing symptoms. The bottom graph shows the posterior probability that the nurse was in the susceptible (blue), infectious (red), and removed (green) states during each day. Nurse 29 refers to one of the nurses {\color{black}who never showed symptoms during the study.} \label{fig:SM_undetected_noro_se} }
\end{figure}

It is also interesting to view plots of $P(S_{it}=s|\bm{y})$ versus $t$ for the nurses {\color{black}who did not show symptoms during the study}. Figure \ref{fig:SM_undetected_noro_se} shows the state estimates for Nurse 29, {\color{black}who is one of the nurses that never showed symptoms.} Note that{\color{black},} due to the homogeneous mixing assumption, the {\color{black}symptomless} nurses are indistinguishable, so the state estimates for any other nurse {\color{black}that did not show symptoms} would be the same as in Figure \ref{fig:SM_undetected_noro_se}. From the figure, {\color{black}we can see that} there is a 25\% chance that a nurse {\color{black}who did not show symptoms during the study} was in the infectious or removed state at the end of the observation period. This implies there should have been around $0.25*61\approx15$ undetected infections. We can more formally estimate the number of undetected infections by looking at the posterior median and 95\% credible interval of $\sum_{i=29}^{89}I[S_{iT}=2 \text{ or }3]$, which is 14 (4, 35). Therefore, there is quite a lot of uncertainty around exactly how many of the nurses {\color{black}were infected but never showed symptoms during the study}.

\clearpage

\subsection{Alternative models} \label{sec:noro_alt}

When analyzing the norovirus outbreak described in Section 4.1 of the main text, \cite{kypraios_tutorial_2017} made the following assumptions:
\begin{enumerate}[noitemsep,nolistsep,label=(\roman*)]
\item That all removal times in the interval $[1,T]$ had been observed and were equal to the symptom-onset times.
\item That there were no undetected infections.
\item That the ward was always open.
\item That there was no background infection risk. That is, they did not include an intercept in their model.
\end{enumerate}

Assumption (i) {\color{black}appears to be} the most commonly made in the literature when analyzing incomplete individual-level epidemic data \citep{oneill_bayesian_1999,o2019markov}. Therefore, in the main text, we focused on investigating the impact of Assumption (i), represented by the known removal times (KRT) model, compared with assuming the removal times had not been observed, represented by the HMM-ILM. Here, we will additionally investigate the consequences of Assumptions (ii)-(iv). 

Assumption (ii) is also common \citep{britton_bayesian_2002,almutiry2021continuous} as, traditionally, it avoids the use of complex reversible jump MCMC steps \citep{oneill_bayesian_1999,jewell_bayesian_2009}. We can incorporate Assumption (ii) by fixing $S_{it}=1$ for all nurses {\color{black}that did not show symptoms}, $i=29,\dots,89$, and $t=0,\dots,9$. We can incorporate Assumption (iii) by setting $W_t=0$ for all $t=0,\dots,9$ in Equation (8) of the main text. Finally, we can incorporate Assumption (iv) by fixing $\alpha=0$ in Equation {\color{black}(2)} of the main text. This means that nurses can only be infected by other nurses; that is, there are no background sources of infection.

Table \ref{tab:SM_tab2} compares the WAIC and posterior summaries of $R_0$ between the HMM-ILM from the main text (which, recall, does not assume the removal times are known), an HMM-ILM with Assumption (i) (the KRT model from the main text), an HMM-ILM with Assumptions (i)+(iii), an HMM-ILM with Assumptions (i)+(iii)+(ii), and an HMM-ILM with Assumptions (i)+(iii)+(ii)+(iv). For any model including Assumption (i), we used the same priors as the KRT model from the main text, $m \sim \text{Unif}(1,3)$, $\alpha \sim \text{Unif}(0,1)$ (if $\alpha$ was included), and $\beta \sim \text{Unif}(0,1)$. In addition, as explained in the main text, for any model including Assumption (i), we replaced Equation {\color{black}(4)} from the main text with Equation (9) from the main text. 

\begin{table}[]
\caption{Shows the WAIC and the posterior median and 95\% CI of $R_0$ for five models fitted to the norovirus outbreak. See Section 4.1 of the main text and Section \ref{sec:noro_alt} for a description of the models.}
\label{tab:SM_tab2}
\begin{tabular}{@{}lcc@{}}
\toprule
\textbf{Model}                                                                                                                             & \textbf{$R_0$ (posterior median and 95\% CI)} & \textbf{WAIC} \\ \midrule
\begin{tabular}[c]{@{}l@{}}Unknown removal times \\[-5pt] (HMM-ILM from main text)\end{tabular}                                                  & 2.61 (1, 7.04)                                & 221           \\
\begin{tabular}[c]{@{}l@{}}Known removal times  \\[-5pt] (KRT model from main text)\end{tabular}                                                  & 0.65 (0.04, 1.82)                              & 222.36        \\
\begin{tabular}[c]{@{}l@{}}Known removal times +  \\[-5pt] Ward always open\end{tabular}                                                          & 0.64 (0.04, 1.81)                              & 222.35        \\
\begin{tabular}[c]{@{}l@{}}Known removal times + \\[-5pt] Ward always open + \\[-5pt] No undetected infections\end{tabular}                              & 0.73 (0.11, 1.52)                              & 222.28        \\
\begin{tabular}[c]{@{}l@{}}Known removal times +  \\[-5pt] Ward always open + \\[-5pt] No undetected infections + \\[-5pt] No intercept\end{tabular} & 1.26 (0.81, 1.92)                             & 222.61        \\ \bottomrule
\end{tabular}
\end{table}

As shown in the table, all models have similar WAIC values despite some models' estimates of $R_0$ being very different. This suggests that there is not enough information in the only 28 {\color{black}symptom-onset times} to distinguish between different epidemic model structures using model comparison criteria. We discuss this issue in more detail in the main text.

From Table \ref{tab:SM_tab2}, the only assumptions that have a large impact on the results are whether the removal times have been observed and whether an intercept is included. We compare the HMM-ILM and KRT models in detail in the main text. Whether to include an intercept depends on whether a nurse could have been infected only by another nurse or whether there were background sources of infection. We would argue for including the intercept, since there were 10 infected patients on the ward during the observation period who could have infected the nurses \citep{caceres1998viral}. As shown in the table, excluding the intercept increases the estimates of $R_0$. This makes sense intuitively. The basic reproduction number measures the amount of infection risk due to between-nurse mixing. If there are no background sources of infection (i.e., no intercept), then all infection risk has to come from between-nurse mixing.

\section{Further Analysis of the Tomato Spotted Wilt Virus (TSWV) Example}

\subsection{Alternative distance kernels}

Here, we compare the following distance kernels for fitting the HMM-ILM to the TSWV experiment in Section 4.2 of the main text:
\begin{enumerate}[noitemsep,nolistsep,label=(\alph*)]
\item Power-law Taylor: $\beta_{j \to i} =\beta_0d_{ij}^{-a}\left(1-\ln(d_{ij})(\beta_1-a)+0.5\ln(d_{ij})^2(\beta_1-a)^2\right)$, where a=1.35 (used in the main text).
\item Power-law exact: $\beta_{j \to i}=\beta_0\beta_1^{-d_{ij}}$.
\item Neighborhood order: $\beta_{j \to i}=\beta_0I[j \in NE_1(i)]+\beta_1I[j \in NE_2(i)]+\beta_2I[j \in NE_3(i)]$, where $NE_k(i)$ denotes the set of all kth order neighbors assuming a queen neighborhood structure \citep{nguyen2025assessing} and, therefore, $NE(i)=NE_1(i)\cup NE_2(i) \cup NE_3(i)$.
\item Linear: $\beta_{j\to i}=\beta_0+\beta_1(3.36-d_{ij})$, where 3.36 meters is the maximum distance between plants $i$ and $j \in NE(i)$.
\item Quadratic: $\beta_{j\to i}=\beta_0+\beta_1(3.36-d_{ij})+\beta_2(3.36-d_{ij})^2$, constrained to be positive (see below).
\item Spline: $\beta_{j\to i}=\beta_0\eta_0(d_{ij})+\beta_{1}\eta_{1}(d_{ij})+\beta_2\eta_{2}(d_{ij})$, where $\eta_k(d_{ij})$ are basis functions (see below).
\end{enumerate}

\begin{figure}[t!]
 	\centering
 	\includegraphics[width=.55\textwidth]{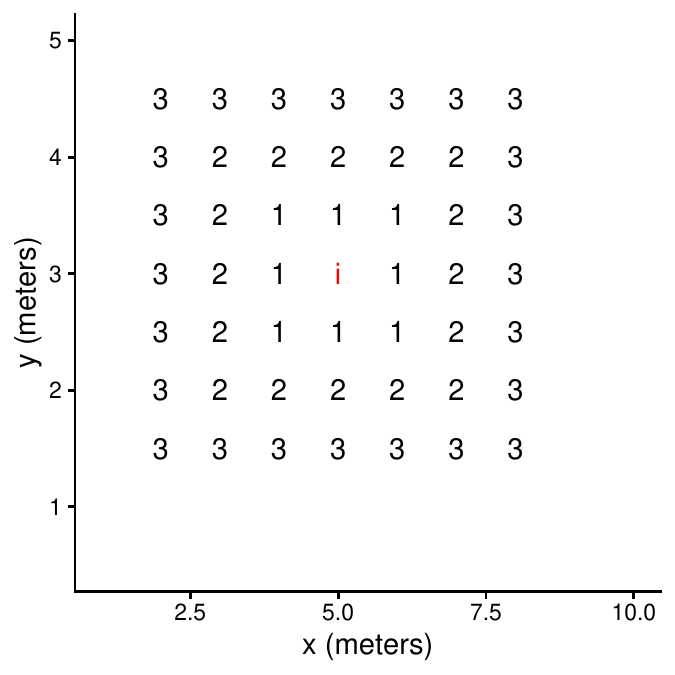}
	\caption{ Shows the first, second, and third order neighbors of plant $i$. Note that the neighborhood set $NE(i)$ used in the main text consists of all first, second, and third-order neighbors (that is, all the plants numbered on the grid). \label{fig:SM_no+plot} }
\end{figure}

For comparing the kernels, we used the same neighborhood set $NE(i)$ as in the main text and the same priors for $\alpha$ and $\theta$. By the kth-order neighbors $NE_k(i)$, we mean as in \cite{nguyen2025assessing}, assuming a queen neighborhood structure; see Figure \ref{fig:SM_no+plot}. The neighborhood set $NE(i)$ used in the main text is equal to $NE_1(i)\cup NE_2(i) \cup NE_3(i)$ and, as can be seen in Figure \ref{fig:SM_no+plot}, consists of all plants at most 3 plants away from plant $i$, including moving diagonally. For the neighborhood order kernel, $\beta_0$ represents the effect of disease spread from first-order neighbors,  $\beta_1$ represents the effect of disease spread from second-order neighbors, and $\beta_2$ represents the effect of disease spread from third-order neighbors. We used wide priors for these effects, $\beta_k \sim \text{Unif}(0,1)$ for $k=0,1,2$.

For the linear kernel, we also used a wide prior for $\beta_k \sim \text{Unif}(0,1)$ for $k=0,1$. To ensure the quadratic kernel is positive on the interval $[0.5, 3.36]$ (0.5 meters is the minimum distance between plants), we imposed the constraints $\beta_0>0$, $\beta_1>0$, and $-\beta_1-2\beta_2(3.36-.5)<0$. These constraints also force the kernel to be strictly decreasing on $[0.5, 3.36]$, which is sensible, as we would expect plants farther away to have a lower effect of disease spread. We used a wide prior for $\beta_0 \sim \text{Unif}(0,1)$ (the effect of disease spread from plants 3.36 meters away), $\beta_1 \sim \text{Unif}(0,20)$, and $\beta_2 \sim N(0,\sigma=5)$. 

For the spline kernel, we used nonnegative natural cubic basis functions constructed using the splines2 package \citep{wang_yan_2021}. We placed a single internal knot at 1.8 meters, which is the average distance between plants $i$ and $j \in NE(i)$. An issue with the spline kernel is that the number of free parameters equals the number of internal knots plus 2. Therefore, using the spline kernel requires estimating a large number of parameters, which is difficult given the amount of missing information in the data. We decided to add only a single internal knot, since the number of free parameters is already at 3, which is one greater than the power-law kernels. We used a wide prior for $\beta_k \sim \text{Unif}(0,5)$ for $k=0,1,2$.

\begin{table}[t!]
\centering
\caption{Shows the WAIC from fitting the HMM-ILM to the TSWV experiment with six different distance kernels. See Section 5.1 for a description of the kernels.}
\label{tab:SM_tab3}
\begin{tabular}{@{}lc@{}}
\toprule
\textbf{Kernel}    & \textbf{WAIC}               \\ \midrule
Power-law Taylor   & 1591.76                     \\
Power-law exact    & 1592.24                     \\
Neighborhood order & 1619.14                     \\
Linear             & 1619.93                     \\
Quadratic          & 1613.29                     \\
Spline             & \multicolumn{1}{l}{1608.22} \\ \bottomrule
\end{tabular}
\end{table}

Table \ref{tab:SM_tab3} compares the WAIC of the six distance kernels. The power-law Taylor kernel has the lowest WAIC, indicating it has the best fit to the data. The difference in WAIC between the power-law Taylor kernel and the non-power-law kernels is significant (greater than 5). The power-law exact kernel has a WAIC very similar to that of the power-law Taylor kernel. This makes sense, as the power-law Taylor kernel is a second-order Taylor approximation of the power-law exact kernel. It also shows that the Taylor approximation does not have a negative effect on the fit of the model. Note that many of the differences in WAIC in Table \ref{tab:SM_tab3} are greater than 5, meaning that we can distinguish fairly well between different distance kernels using WAIC at this sample size (327 {\color{black}symptom-onset times}). In particular, moving from the linear kernel to the quadratic kernel, then to the spline kernel, and finally to the power-law kernels, all show a significant decrease in WAIC.

\begin{figure}[t!]
 	\centering
 	\includegraphics[width=\textwidth]{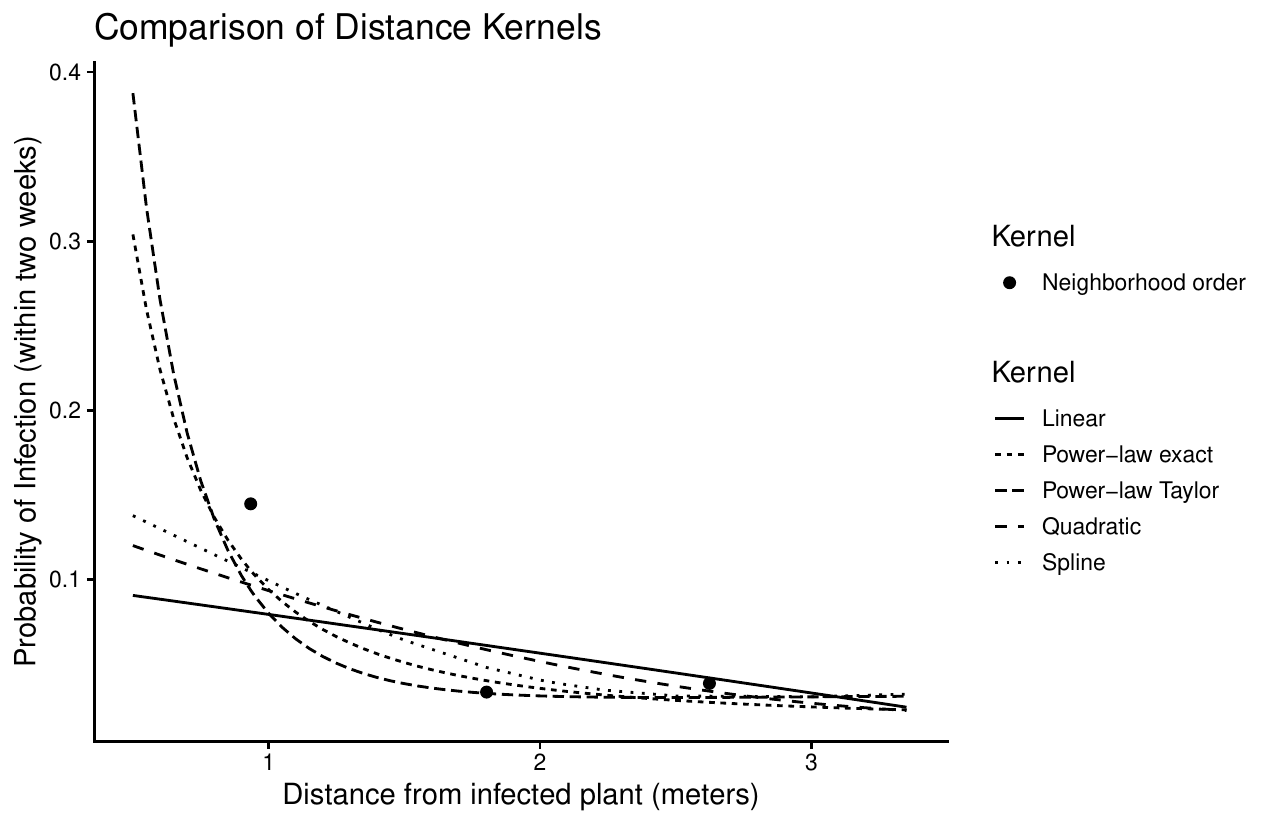}
	\caption{Shows the probability that a
susceptible plant is infected (within two weeks), given a single infectious plant, as a function
of the distance between the two plants, for six distance kernels. Shows the posterior medians. Note that the neighborhood order kernel is not a function of distance but of neighborhood order (1,2,3). Therefore, we placed the points at the average distance between plants $i$ and $j \in NE_k(i)$ for $k=1,2,3$. \label{fig:SM_dk_compare} }
\end{figure}

Figure \ref{fig:SM_dk_compare} shows the probability of infection for a susceptible plant versus distance from an infectious plant for the six distance kernels. The power-law kernels estimate a much higher risk of infection for a susceptible plant that is half a meter away (one above or below in the same row) from an infectious plant. However, all kernels agree that there is little infection risk beyond 3 meters, indicating that the disease has a short range of dispersal. Overall, the figure shows that the kernels eventually converge, but that estimates of the effect of disease spread between close plants can be very sensitive to the choice of kernel.

\subsection{Posterior distributions}

\begin{figure}[h]
 	\centering
 	\includegraphics[width=\textwidth]{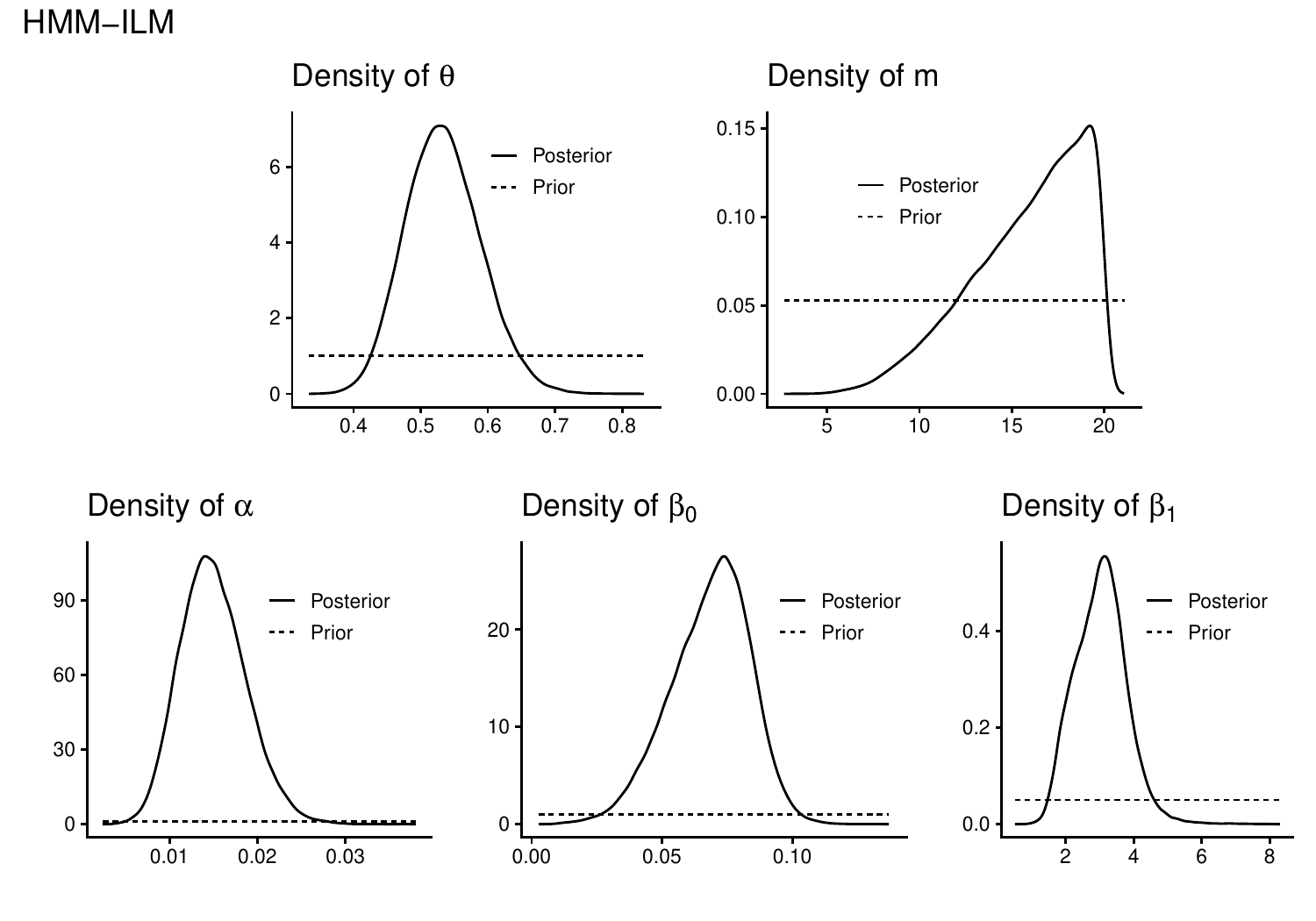}
	\caption{Shows the posterior and prior distributions of $\theta$, $m$, $\alpha$, $\beta_0$, and $\beta_1$ from fitting the
HMM-ILM from Section 4.2 of the main text to the TSWV experiment. \label{fig:SM_postHMM_ILM} }
\end{figure}

\begin{figure}[h]
 	\centering
 	\includegraphics[width=\textwidth]{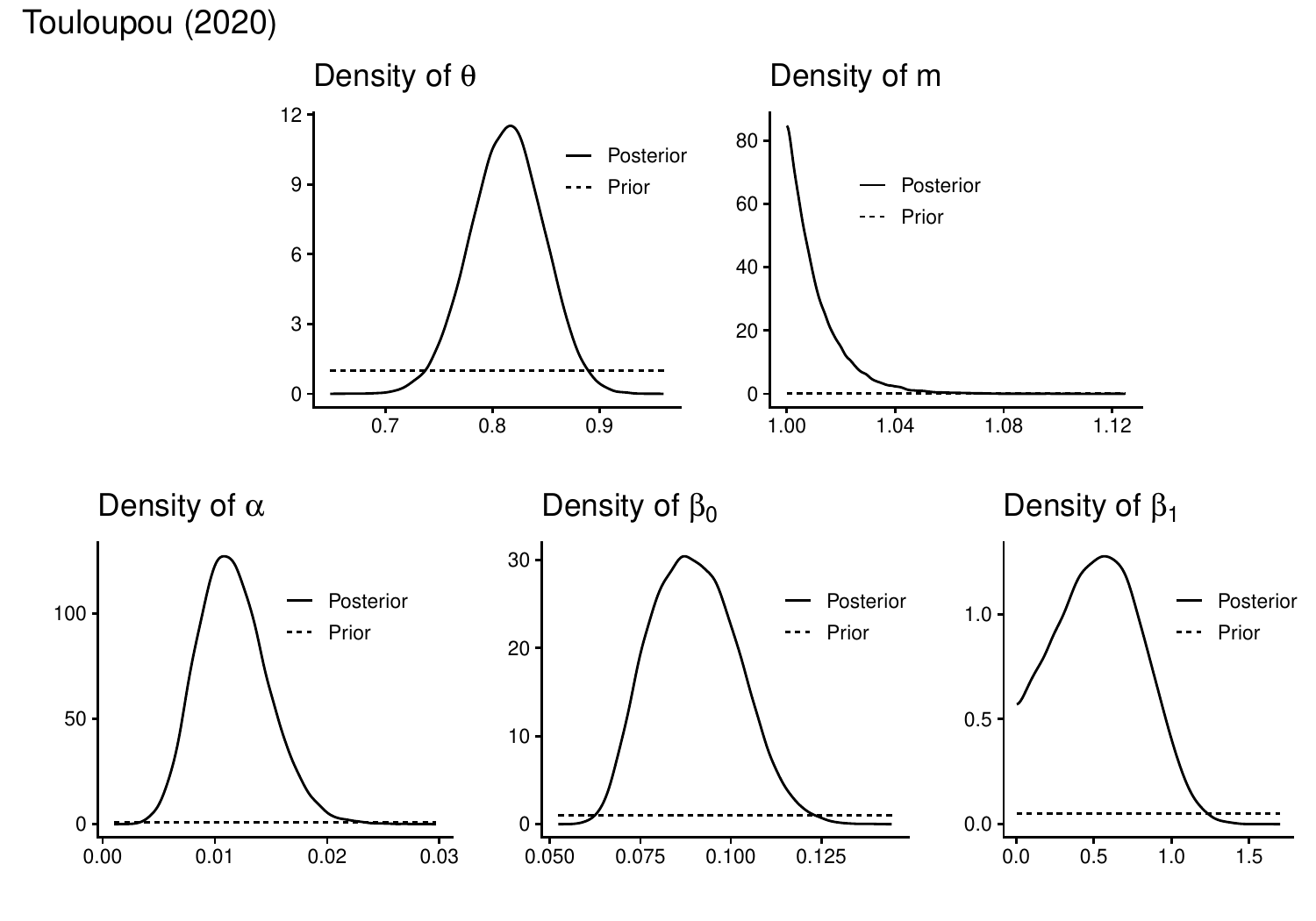}
	\caption{Shows the posterior and prior distributions of $\theta$, $m$, $\alpha$, $\beta_0$, and $\beta_1$ from fitting the
Touloupou (2020) model from Section 4.2 of the main text to the TSWV experiment. \label{fig:SM_post_touloupou} }
\end{figure}

\begin{figure}[h]
 	\centering
 	\includegraphics[width=\textwidth]{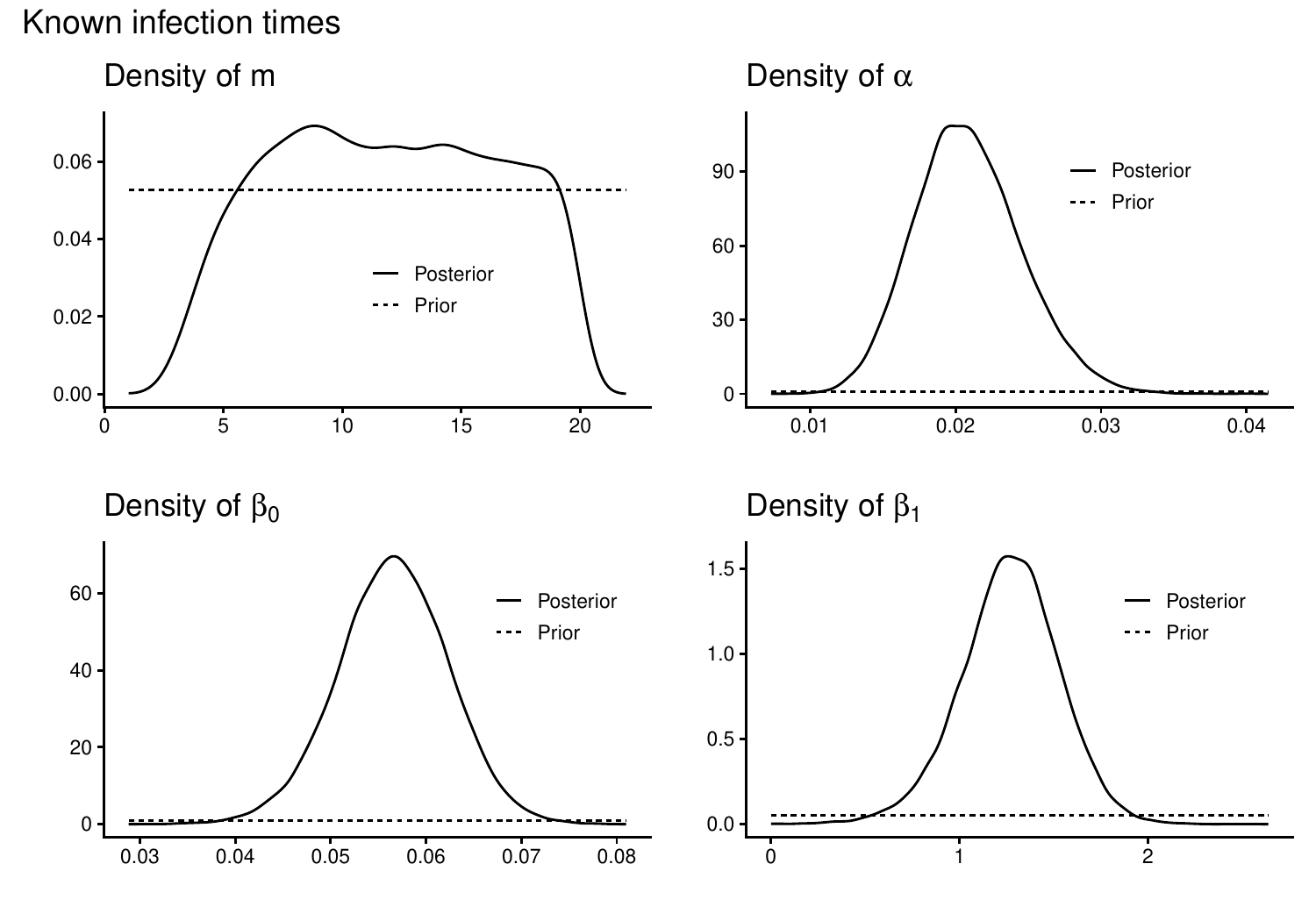}
	\caption{Shows the posterior and prior distributions of $m$, $\alpha$, $\beta_0$, and $\beta_1$ from fitting the
model assuming known infection times from Section 4.2 of the main text to the TSWV experiment. \label{fig:SM_post_KIT} }
\end{figure}

\begin{figure}[h]
 	\centering
 	\includegraphics[width=\textwidth]{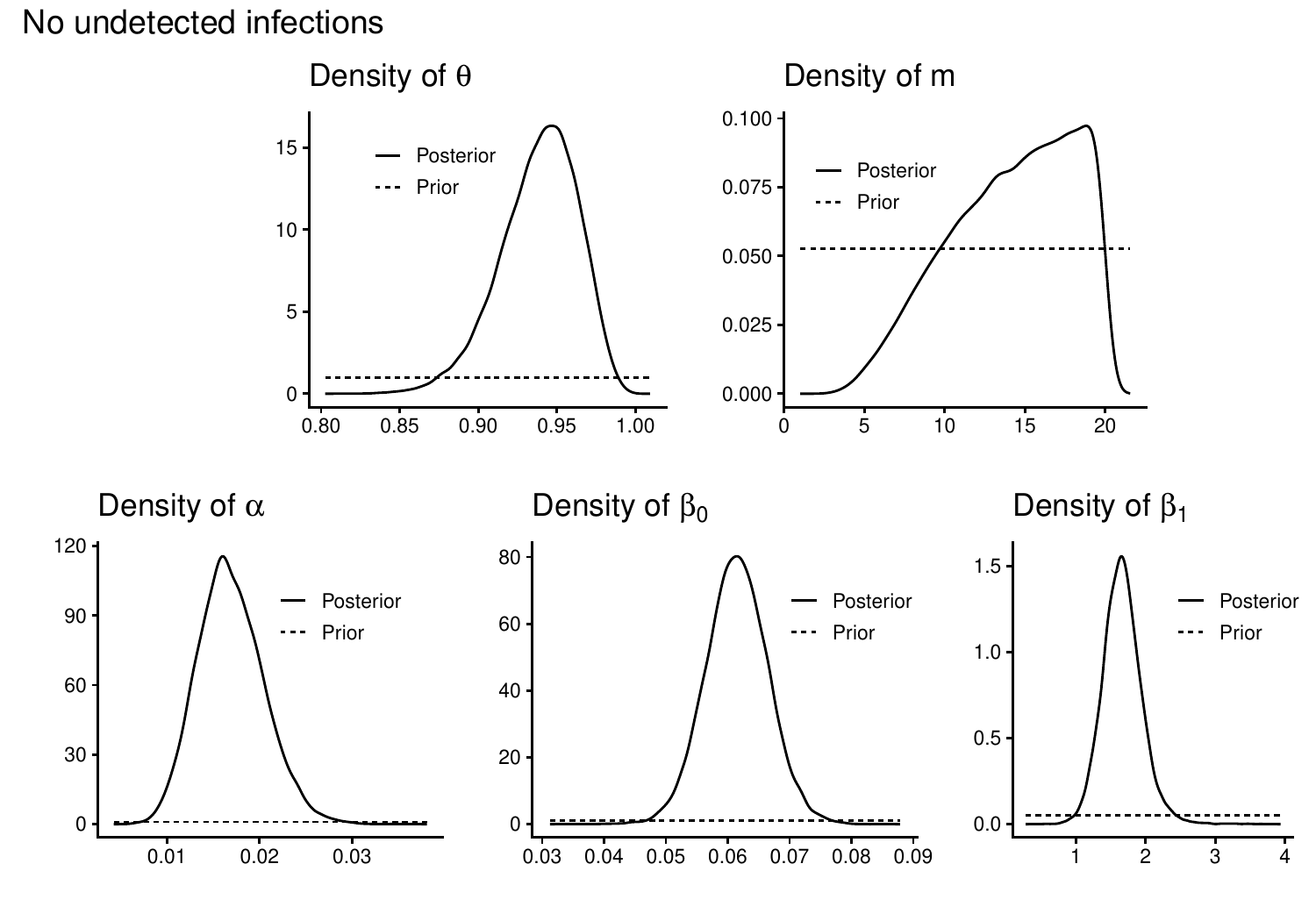}
	\caption{Shows the posterior and prior distributions of $\theta$, $m$, $\alpha$, $\beta_0$, and $\beta_1$ from fitting the
model assuming that {\color{black}all plants that did not show symptoms escaped infection} (from Section 4.2 of the main text) to the TSWV experiment. \label{fig:SM_nui_post} }
\end{figure}

Here, we present the complete set of posterior distributions for the models shown in Figure {\color{black}8} of the main text. Figure \ref{fig:SM_postHMM_ILM} shows the posteriors for the HMM-ILM, Figure \ref{fig:SM_post_touloupou} shows the posteriors for the Touloupou (2020) model, Figure \ref{fig:SM_post_KIT} shows the posteriors for the model assuming known infection times, and  Figure \ref{fig:SM_nui_post} shows the posteriors of the model assuming no undetected infections. See Section 4.2 of the main text for a description of each model.

\vfill
\vfill

\clearpage
\bibliography{HMMILM}